\documentclass[12pt]{article} \usepackage{epsf} \usepackage{epsfig}
\newcommand{\lsim}{
\mathrel{\hbox{\rlap{\hbox{\lower4pt\hbox{$\sim$}}}\hbox{$<$}}}}

\newcommand{\gsim}{
\mathrel{\hbox{\rlap{\hbox{\lower4pt\hbox{$\sim$}}}\hbox{$>$}}}}

\begin{document}
\begin{titlepage}
\vspace*{-0.5truecm}

\begin{flushright}
IPPP/06/23\\ DCPT/06/46\\ CERN--PH--TH/2006--065\\ hep-ph/0604249
\end{flushright}

\vspace*{2.0truecm}

\begin{center}
\boldmath {\Large{\bf Probing New Physics through $B$
Mixing:\\[3pt] Status, Benchmarks and Prospects}}
\end{center}

\vspace{0.7truecm}

\begin{center}
{\bf Patricia Ball${}^{a,b}$ and Robert Fleischer${}^b$}
 
\vspace{0.4truecm}

${}^a$ {\sl IPPP, Department of Physics, University of Durham, Durham
DH1 3LE, UK}

\vspace{0.2truecm}

${}^b$ {\sl Theory Division, Department of Physics, CERN, CH-1211
Geneva 23, Switzerland}

\end{center}

\vspace{0.9cm}
\begin{abstract}
\vspace{0.2cm}\noindent As is well known, $B^0_{d,s}$--$\bar
B^0_{d,s}$ mixing offers a profound probe into the effects of physics
beyond the Standard Model. The data obtained at the $e^+e^-$ $B$
factories have already provided valuable insights into the $B_d$-meson
system, and very recently also the $B^0_s$--$\bar B^0_s$ oscillation
frequency $\Delta M_s$ has been measured at the Tevatron. We give a
critical discussion of the interpretation of these data in terms of
model-independent new-physics parameters.  We address in particular
the impact of the uncertainties of the relevant input parameters, set
benchmarks for their accuracies as required by future precision
measurements at the LHC, and explore the prospects for new
CP-violating effects in the $B_s$ system. To complement our
model-independent analysis, we also discuss the constraints imposed by
the CDF measurement of $\Delta M_s$ on popular models of new physics,
namely scenarios with an extra $Z'$ boson
and supersymmetry.  We find that the new data still leave sizeable room for 
new-physics contributions to $B^0_{s}$--$\bar B^0_{s}$ mixing, which could be 
detected at the LHC.
\end{abstract}

\vfill


\vspace*{0.5truecm}
\vfill
\noindent
April 2006

\end{titlepage}

\thispagestyle{empty}
\vbox{}
\newpage

\setcounter{page}{1}

\section{Introduction}
One of the most promising ways to detect the effects of new physics (NP) on 
$B$ decays is to look for deviations of flavour-changing neutral-current 
(FCNC) processes from their Standard Model (SM) predictions;
FCNC processes only occur at the loop-level in the SM and
hence are particularly sensitive to NP virtual particles and
interactions. A prominent example that has received extensive
experimental and theoretical attention is $B^0_q$--$\bar B^0_q$ mixing
($q\in\{d,s\}$), which, in the SM, is due to box diagrams with $W$-boson
and up-type quark exchange. In the language of effective field theory, these diagrams
induce an effective Hamiltonian, which causes $B^0_q$ and $\bar B^0_q$ 
mesons to mix and generates a $\Delta B=2$ transition:
\begin{equation}\label{eq1}
\langle B_q^0| {\cal H}^{\Delta B=2}_{\rm eff} | \bar B_q^0\rangle = 2 M_{B_q}
M_{12}^q\,,
\end{equation}
where $M_{B_q}$ is the $B_q$-meson mass. Thanks to $B^0_q$--$\bar B^0_q$
mixing, an initially present $B^0_q$ state evolves into a time-dependent linear
combination of $B^0_q$ and $\bar B^0_q$ flavour states. The oscillation frequency
of this phenomenon is characterized by the mass difference of the ``heavy"
and ``light" mass eigenstates,
\begin{equation}\label{DM-def}
\Delta M_q\equiv M_{\rm H}^{q}-M_{\rm L}^{q} = 2 |M_{12}^q|\,,
\end{equation}
and the CP-violating mixing phase 
\begin{equation}\label{phiq-def}
\phi_q = \arg M_{12}^q\,,
\end{equation}
which enters ``mixing-induced" CP violation. The mass difference $\Delta M_q$ can be -- 
and has been -- measured from the proper-time distribution of $B^0_q$ candidates
identified through their decays into (mostly) flavour-specific modes, after having been 
tagged as mixed or unmixed. The current experimental results are
\begin{equation}\label{exp}
\Delta M_d = (0.507\pm 0.004)\,{\rm ps}^{-1}\,,
\qquad \Delta M_s = \left[17.33^{+0.42}_{-0.21}({\rm
  stat})\pm 0.07({\rm syst})\right]\,{\rm ps}^{-1}\,,
\end{equation}
where the value of $\Delta M_d$ is the world average quoted by the
``Heavy Flavour Averaging Group" (HFAG) \cite{HFAG}. Concerning
$\Delta M_s$, only lower bounds were available for many years from the 
LEP experiments at CERN and SLD at SLAC \cite{LEPBOSC}. Since the 
currently operating $e^+e^-$ $B$ factories run at the $\Upsilon(4S)$ resonance, 
which decays  into $B_{u,d}$, but not into $B_s$ mesons, the $B_s$ system 
cannot be explored by the BaBar and Belle experiments. However, plenty of 
$B_s$ mesons are produced at the Tevatron (and later on will be at the LHC), which --
very recently -- allowed the CDF collaboration to measure $\Delta M_s$ with the 
result given above \cite{CDF}; the D0 collaboration has provided, also
very recently, a two-sided bound on $\Delta M_s$ at the 90\% C.L. \cite{D0}:
\begin{equation}\label{D0-bound}
17 \,{\rm ps}^{-1}< \Delta M_s < 21\,{\rm ps}^{-1},
\end{equation}
which is compatible with the CDF measurement and corresponds to a
$2.5\,\sigma$ signal at $\Delta M_s=19\,{\rm ps}^{-1}$. These new results
from the Tevatron have already triggered a couple of phenomenological
papers \cite{CMNSW}--\cite{FOR}.

In the SM, $M_{12}^q$ is given by
\begin{equation}\label{M12SM}
M_{12}^{q,{\rm SM}} = \frac{G_{\rm F}^2M_W^2}{12\pi^2}M_{B_q}\hat{\eta}^{B}
\hat B_{B_q}f_{B_q}^2(V_{tq}^\ast V_{tb})^2 S_0(x_t)\,,
\end{equation}
where $G_{\rm F}$ is Fermi's constant, $M_W$ the mass of the $W$ boson, 
$\hat{\eta}^{B}=0.552$ a short-distance QCD correction (which is the same for
the $B^0_d$ and $B^0_s$ systems) \cite{BBL}, whereas
the ``bag" parameter
$\hat B_{B_q}$ and the decay constant $f_{B_q}$ are non-perturbative quantities.
$V_{tq}$ and $V_{tb}$ are elements of the 
Cabibbo--Kobayashi--Maskawa (CKM) matrix \cite{cab,KM}, and 
$S_0(x_t\equiv \overline{m}_t^2/M_W^2)=2.35\pm0.06$ with $\overline{m}_t(m_t) =
(164.7\pm 2.8)\,{\rm GeV}$ \cite{top} is one of the ``Inami--Lim" functions \cite{IL},
describing the $t$-quark mass dependence of the box diagram with internal 
$t$-quark exchange; the contributions of internal $c$ and $u$ quarks
are, by virtue of the Glashow--Iliopoulos--Maiani (GIM) mechanism \cite{GIM},
suppressed by  $(m_{u,c}/M_W)^2$. 

The mixing phases $\phi_q$ can be measured from ``mixing-induced" CP
asymmetries. 
In the SM, one has
\begin{equation}\label{phi-SM}
\phi_d^{\rm SM} = 2\beta\,,\qquad
\phi_s^{\rm SM} = -2 \delta\gamma\,,
\end{equation}
where $\beta$ is the usual angle of the ``conventional" unitarity triangle (UT)
of the CKM matrix, while $\delta\gamma$ characterizes another unitarity triangle 
\cite{LHC-report} that differs from the UT through ${\cal O}(\lambda^2)$ 
corrections in the Wolfenstein expansion \cite{wolf}.\footnote{Throughout this paper,
we use the phase convention for the CKM matrix advocated by the Particle 
Data Group \cite{PDG}, where the decay amplitudes of $b\to c\bar c s$
processes carry essentially no CP-violating weak phase. Physical CP asymmetries
are of course independent of the applied CKM phase convention, as shown 
explicitly in Ref.~\cite{RF-habil}.}

The purpose of this paper is to explore the possibility that $B^0_q$--$\bar B^0_q$
mixing is modified by NP contributions at the tree level and/or new particles in the 
loops. We shall find in particular that -- despite the apparently strong constraints 
posed by the precise measurements of $\Delta M_q$ in Eq.~(\ref{exp}) -- these 
results can contain potentially large NP contributions, which presently cannot 
be detected. 

The outline of our paper is as follows: in Section~\ref{sec:input}, we
collect the input parameters of our analysis and discuss the status
of the relevant hadronic uncertainties. In Sections~\ref{sec:Bd} and 
\ref{sec:Bs}, we then focus on the $B_d$- and $B_s$-meson systems, 
respectively, and investigate, in a model-independent way, the size of
possible NP contributions to $\Delta M_q$ and $\phi_q$ in the
light of present and future experimental measurements and hadronic
uncertainties. In this analysis, we consider also a scenario for the
experimental and theoretical situation in the year 2010, and set
benchmarks for the required accuracy of the relevant hadronic parameters. 
It turns out that the situation in the $B_s$ system 
is more favourable than in the $B_d$ system, and that still ample space for 
NP effects in $B^0_s$--$\bar B^0_s$ mixing is left, which could be detected 
at the LHC.
In Section~\ref{sec:NP-models}, we complement the model-independent
discussion of Sections~\ref{sec:Bd} and \ref{sec:Bs} by analyses of two
specific scenarios for NP: models with an extra $Z'$ boson 
and supersymmetry (SUSY)  with an approximate alignment of quark and
squark masses. We summarize our conclusions in Section~\ref{sec:concl}.

\boldmath
\section{Input Parameters and Hadronic Uncertainties}\label{sec:input}
\unboldmath
\boldmath
\subsection{CKM Parameters}
\unboldmath
Before going into the details of $B^0_q$--$\bar B^0_q$ mixing and possible
NP effects, let us first have a closer look at the relevant input parameters and
their uncertainties. Throughout our analysis, we assume that the CKM matrix
is unitary, and shall use this feature to express the CKM elements entering
$B^0_q$--$\bar B^0_q$ mixing in terms of quantities that can be determined
through tree-level processes of the SM. The key r\^ole is then played by 
$|V_{cb}|$ and $|V_{ub}|$. The former quantity is presently known with 
2\% precision from semileptonic $B$ decays; in this paper we shall use the 
value obtained in Ref.~\cite{Buchmuller} from the analysis of leptonic and 
hadronic moments in inclusive $b\to c \ell \bar\nu_\ell$ transitions \cite{Gambino}:
\begin{equation}\label{Vcb}
|V_{cb}| = (42.0\pm 0.7)\cdot 10^{-3}\,;
\end{equation}
this value agrees with that from exclusive decays. 

The situation is
less favourable with $|V_{ub}|$: there is a 1$\,\sigma$ discrepancy
between the values from inclusive and exclusive $b\to u\ell\bar\nu_\ell$
transitions \cite{HFAG}:
\begin{equation}
|V_{ub}|_{\rm incl} = (4.4\pm 0.3)\cdot 10^{-3}\,,\qquad 
|V_{ub}|_{\rm excl} = (3.8\pm 0.6)\cdot 10^{-3}\,.
\end{equation}
The error on $|V_{ub}|_{\rm excl}$ is dominated by the theoretical
uncertainty of lattice and light-cone sum rule calculations of $B\to\pi$ and
$B\to\rho$ transition form factors \cite{Vublatt,LCSR}, whereas for
$|V_{ub}|_{\rm incl}$ experimental and theoretical errors are at par.
We will use both results in our analysis.

Whereas any improvement of the error of $|V_{cb}|$ will have only
marginal impact on the analysis of $B$ mixing, a reduction of the
uncertainty of $|V_{ub}|$ will be very relevant. As a benchmark
scenario for the situation in 2010, we will assume that the central
value of $|V_{ub}|_{\rm incl}$ gets confirmed and that its
uncertainty will shrink to $\pm 0.2\cdot 10^{-3}$, i.e.\ 5\%, thanks to
better statistics and an increased precision of theoretical
predictions, for instance from further developments in the dressed gluon
exponentiation \cite{einan}.

\boldmath
\subsection{Hadronic Mixing Parameters 
$f_{B_q}\hat{B}_{B_q}^{1/2}$}\label{ssec:hadr-uncert}
\unboldmath
The next ingredient in the SM prediction for $M_{12}^{q, {\rm SM}}$ are
the non-perturbative matrix elements $f_{B_q}^2 \hat{B}_{B_q}$.
These parameters have been the subject of numerous lattice calculations,
both quenched and unquenched,  using various lattice actions and  
implementations of both heavy and light quarks. The current front runners are unquenched
calculations with 2 and 3 dynamical quarks, respectively, and Wilson
or staggered light quarks. Despite tremendous progress in recent
years, the results still suffer from a variety of uncertainties which
is important to keep in mind when interpreting and using lattice
results. 
One particular difficulty in determining
$f_{B_d}$ is the chiral extrapolation needed to go to the physical
$d$-quark mass.\footnote{Many lattice simulations do not
  distinguish between $u$- and $d$-quark masses and use $m_{u, d} \equiv
  (m_u+m_d)/2$. The physical value of the light-quark mass ratio is then
  $m_{u, d}/m_s=0.041\pm 0.003$ from chiral perturbation theory
  \cite{Leutwyler}.} 
Lattice calculations are usually performed at
unphysically large $u$- and $d$-quark masses, as the simulation of
dynamical fermions involves many inversions of the fermions'
functional determinant in the path integral and is very dear in terms
of CPU time. Therefore, an extrapolation, called the chiral
extrapolation, in the light-quark masses from feasible to
physical  masses is necessary, which is done
using the functional form predicted by chiral
perturbation theory. Based on these arguments, the
chiral extrapolation of $\hat B_{B_d}$ to the physical limit is
expected to be smooth, 
whereas that of $f_{B_d}$ is potentially prone to logarithms
\cite{sinead}, which leads to a considerable increase in the
uncertainty. The most recent (unquenched) simulation by the JLQCD collaboration
\cite{JLQCD}, with non-relativistic $b$ quarks and two flavours of
dynamical light (Wilson) quarks, yields $f_{B_d} = (0.191\pm
0.010^{+0.012}_{-0.022})\,$GeV and
\begin{eqnarray}
\left.f_{B_d}\hat{B}_{B_d}^{1/2}\right|_{\rm JLQCD} &=& (0.215\pm
0.019^{+0}_{-0.023})\,{\rm GeV}\,,\nonumber\\
\left.f_{B_s}\hat{B}_{B_s}^{1/2}\right|_{\rm JLQCD} &=& (0.245\pm
0.021^{+0.003}_{-0.002})\,{\rm GeV}\,,\nonumber\\
\xi_{\rm JLQCD} \equiv 
\left.\frac{f_{B_s}\hat{B}_{B_s}^{1/2}}{f_{B_d}\hat{B}_{B_d}^{1/2}}
\right|_{\rm JLQCD}& = & 1.14\pm 0.06^{+0.13}_{-0}\,,\label{JLQCD}
\end{eqnarray}
where the first error includes uncertainties from statistics and
various systematics, where\-as the second, asymmetric error comes
from the chiral extrapolation. Note that part of the systematic errors
cancel in the ratio $\xi$. In this calculation, the ratio
$m_{u, d}/m_s$ was varied between 0.7 and 2.9. 

More recently, (unquenched) simulations
with three dynamical flavours have become possible using staggered
quark actions. The HPQCD collaboration obtains $f_{B_d} = (0.216\pm
0.022)\,$MeV \cite{HPQCD}, where a ratio of $m_{u, d}/m_s$ as small as 0.125 could be
achieved, due to the good chiral properties of the staggered action.
This implies that the chiral extrapolation is less critical and the corresponding error much
smaller. The quoted error on $f_{B_q}$ is now dominated by yet
uncalculated higher-order matching terms which are needed to
match the (effective theory) lattice calculations to continuum QCD.
Lacking any direct calculation of $\hat B_{B_q}$ with three dynamical
flavours, and in view of the fact that the bag parameter is likely to be less
sensitive to chiral extrapolation, it has been suggested to combine
the results of $f_{B_q}$ from HPQCD with that of $\hat B_{B_q}$ from
JLQCD, yielding \cite{Okamoto}:
\begin{eqnarray}
\left.f_{B_d}\hat{B}_{B_d}^{1/2}\right|_{\rm (HP+JL)QCD}& =& (0.244\pm
0.026)\,{\rm GeV}\,,\nonumber\\
\left.f_{B_s}\hat{B}_{B_s}^{1/2}\right|_{\rm (HP+JL)QCD} &=&
(0.295\pm0.036)\,{\rm GeV}\,,\nonumber\\
\xi_{\rm (HP+JL)QCD} & = &
1.210^{+0.047}_{-0.035}\,,\label{HPQCD}
\end{eqnarray}
where all errors are added in quadrature. 

Although we shall use both (\ref{JLQCD}) and (\ref{HPQCD}) in our
analysis, we would like to stress that the
errors are likely to be optimistic. Apart from the issue of the 
chiral extrapolation discussed above, there is also the question of
discretisation effects (JLQCD uses data obtained at only one lattice
spacing) and the renormalisation of matrix elements (for lattice
actions without chiral symmetry, the axial vector current is not
conserved and $f_{B_q}$ needs to be renormalised), which some argue should
be done in a non-perturbative way \cite{alpha}. Simulations with
staggered quarks also face potential problems with unitarity,
locality and an odd number of flavours (see, for instance,
Ref.~\cite{staggered}). A confirmation of the HPQCD results by
simulations using the (theoretically better understood) Wilson action
with small quark masses will certainly be highly welcome.

Given this situation, we consider it not very likely that the errors on
$f_{B_q}$, $\hat{B}_{B_q}$ and $\xi$ will come down considerably  in the near
future. For our benchmark 2010 scenario, we hence will assume the
values of hadronic parameters and uncertainties given in (\ref{HPQCD}).

We are now well prepared for the discussion of $B^0_q$--$\bar B^0_q$
mixing.

\boldmath
\section{The $B_d$-Meson System}\label{sec:Bd}
\unboldmath
\subsection{Model-Independent NP Parameters}
\begin{figure}[t]
\centerline{
\epsfxsize=0.47\textwidth\epsffile{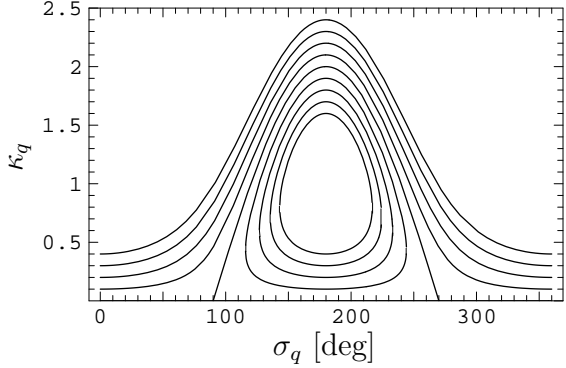}
 }
 \vspace*{-0.3truecm}
\caption[]{The dependence of $\kappa_q$ on $\sigma_q$ for values of 
$\rho_q$ varied between 1.4 (most upper curve) and 0.6 (most inner curve),
in steps of 0.1. The contours apply to both the $q=d$ and the $q=s$
system.}\label{fig:kappa-rho-d}
\end{figure}

Let us first have a closer look at the $B^0_d$--$\bar B^0_d$ mixing
parameters.
In the presence of NP, the matrix element $M_{12}^d$ can be
written, in a model-independent way, as
$$M_{12}^d = M_{12}^{d,{\rm SM}} \left(1 + \kappa_d e^{i\sigma_d}\right)\,,$$
where the real parameter $\kappa_d\geq 0$ measures the ``strength'' of
the NP contribution with respect to the SM, whereas $\sigma_d$ is a new CP-violating
phase; analogous formulae apply to the $B_s$ system. 
The $B_d$ mixing parameters then read
\begin{eqnarray}
\Delta M_d & = & \Delta M_d^{\rm SM}\left[ 1 + \kappa_d
  e^{i\sigma_d}\right],\label{M-d}\\
\phi_d & = & \phi_d^{\rm SM}+\phi_d^{\rm NP}=
\phi_d^{\rm SM} + \arg (1+\kappa_d e^{i\sigma_d})\,.\label{phi-d}
\end{eqnarray}
The experimental result for $\Delta M_d$ and the theoretical
prediction $ \Delta M_d^{\rm SM}$ provide the following constraint on
$\kappa_d$ and $\sigma_d$:
\begin{equation}\label{rhod-def}
\rho_d\equiv
\left|\frac{\Delta M_d}{\Delta M_d^{\rm SM}}\right|=
\sqrt{1+2\kappa_d\cos\sigma_d+\kappa_d^2}\,,
\end{equation}
which determines, for instance,
$\kappa_d$ as function of $\sigma_d$:
\begin{equation}\label{kappa-sig-rho}
\kappa_d=-\cos\sigma_d\pm\sqrt{\rho_d^2-\sin^2\sigma_d}\,.
\end{equation}
In Fig.~\ref{fig:kappa-rho-d}, we illustrate the corresponding contours in the
$\sigma_d$--$\kappa_d$ plane for values of $\rho_d$ between 0.6 and
1.4, varied in steps of 0.1.
Interestingly enough, a value of
$\rho_d$ smaller than 1 imposes a constraint on the weak NP phase
$\sigma_d$:
\begin{equation}\label{sigma-bound}
\pi-\arcsin\rho_d\leq\sigma_d\leq\pi+\arcsin\rho_d. 
\end{equation}

\boldmath
\subsection{The SM Prediction for $\Delta M_d$}\label{ssec:DMd-SM}
\unboldmath
In order to make use of these constraints, one needs to know
the SM prediction $\Delta M_d^{\rm SM}$. In particular, one has to make sure that
the parameters entering $M_{12}^{d,{\rm SM}}$, Eq.~(\ref{M12SM}), are free from
NP. This can be achieved, to very good accuracy, by expressing the relevant 
CKM factor in $\Delta M_d^{\rm SM}$ in terms of parameters measured in tree-level
processes. To this end, we use the Wolfenstein parametrization \cite{wolf},
as generalized in Ref.~\cite{blo}, and the unitarity of the CKM matrix to write
\begin{equation}\label{CKM-Bd}
|V_{td}^\ast V_{tb}|=|V_{cb}|\lambda\sqrt{1-2R_b\cos\gamma+R_b^2}\,.
\end{equation}
Here the quantity $R_b$ is given by
\begin{equation}\label{Rb-def}
R_b\equiv\left(1-\frac{\lambda^2}{2}\right)\frac{1}{\lambda}\left|\frac{V_{ub}}{V_{cb
}}\right|
=\sqrt{\bar\rho^2+\bar\eta^2},
\end{equation}
with
\begin{equation}\label{rho-bar-eta-bar}
\bar\rho=(1-\lambda^2/2)\rho=R_b\cos\gamma, \quad
\bar\eta=(1-\lambda^2/2)\eta=R_b\sin\gamma\,;
\end{equation}
$R_b$ measures one side of the UT, and $\gamma$ denotes the usual UT angle. 

As we saw in Section~\ref{sec:input}, $|V_{cb}|$ and $|V_{ub}|$ 
can be determined from semileptonic $B$ decays, which arise at tree
level in the SM and hence are  very robust with respect to NP effects. A
similar comment applies to the Wolfenstein parameter $\lambda\equiv|V_{us}|$
\cite{wolf,blo}, which can be determined, for instance, 
from $K\to\pi\ell\bar\nu_\ell$ decays.
Using the values of $|V_{cb}|$ and $|V_{ub}|$ discussed in Section~\ref{sec:input}
and $\lambda =0.225\pm0.001$ \cite{blucher}, we obtain
\begin{equation}\label{Rb}
R_b^{\rm incl} = 0.45\pm 0.03\,,\qquad R_b^{\rm excl} = 0.39\pm
0.06\,,
\end{equation}
where the labels ``incl" and ``excl" refer to the determinations of $|V_{ub}|$
through inclusive and exclusive $b\to u\ell\bar \nu_\ell$ transitions, 
respectively.

The angle $\gamma$ can be determined in a variety of ways 
through CP-violating effects in pure 
tree decays of type $B\to D^{(*)} K^{(*)}$ \cite{WG5-report}. Using the
present $B$-factory data, the following results were obtained through a 
combination of various methods:
\begin{equation}\label{gam-DK}
\left.\gamma\right|_{D^{(*)} K^{(*)}} = \left\{
\begin{array}{ll}
(62^{+35}_{-25})^\circ & \mbox{(CKMfitter collaboration
    \cite{CKMfitter}),}\\[5pt] 
(65\pm 20)^\circ & \mbox{(UTfit collaboration \cite{UTfit})}.
\end{array}
\right.
\end{equation}
A more precise value for $\gamma$ was obtained in Ref.~\cite{BFRS-05},
from the $B$-factory data on CP asymmetries in 
$B^0_d\to\pi^+\pi^-$ and $B^0_d\to\pi^-K^+$ decays, which receive both
tree and
penguin contributions:
\begin{equation}\label{gam-piK}
\left.\gamma\right|_{\pi^+\pi^-,\pi^-K^+} = (73.9^{+5.8}_{-6.5})^\circ.
\end{equation}
Within the NP scenario of modified electroweak penguins considered
in Ref.~\cite{BFRS-05}, (\ref{gam-piK}) is not affected by NP
effects. The central value of (\ref{gam-piK}) is higher than that of 
(\ref{gam-DK}), but both results are perfectly consistent
because of the large errors of the $B\to D^{(*)} K^{(*)}$ determinations. 
An even larger value of $\gamma$ in the ballpark of $80^\circ$ was 
recently extracted from  $B\to\pi\pi$ data with the help of  ``soft collinear
effective theory" (SCET) \cite{SCET}. 

In our analysis,  we use the UTfit value
\begin{equation}\label{gamma}
\gamma=(65\pm 20)^\circ.
\end{equation}
With the help of (\ref{Vcb}), (\ref{CKM-Bd}) and (\ref{Rb}), we then obtain
\begin{equation}
|V_{td}^* V_{tb}|_{\rm incl} = (8.6\pm 1.5)\cdot 10^{-3}\,,\qquad 
|V_{td}^* V_{tb}|_{\rm excl} = (8.6\pm 1.3)\cdot 10^{-3}\,,
\end{equation}
where the uncertainty is dominated by that of the angle $\gamma$. 

For our 2010 benchmark scenario, we assume that the central value
of $\gamma$ will settle at $70^\circ$, and that the error will shrink to $\pm 5^\circ$ 
thanks to strategies using pure tree decays of $B_{u,d}$ and $B_s$ mesons 
for the determination of $\gamma$, which can be implemented at the LHC. In fact,
a statistical accuracy of $\sigma_{\rm stat}(\gamma)\approx 2.5^\circ$ is expected
at LHCb after 5 years of taking data \cite{schneider}.

For the convenience of the reader, we summarise all CKM input parameters, as well 
as their counterparts for the $B_s$ system to be discussed in Section~\ref{sec:Bs},  
in Tab.~\ref{tab:CKM}; in Tab.~\ref{tab:2010}, we give the input data for our 2010 scenario.

\begin{table}[tbp]
\renewcommand{\arraystretch}{1.2}
\addtolength{\arraycolsep}{3pt}
$$
\begin{array}{|l||c|l|l|}
\hline
{\rm Parameter} & {\rm Value} & {\rm Ref.} & {\rm Remarks}\\\hline
\lambda & 0.225\pm0.001 & \cite{blucher} & \mbox{CKM05 average}\\
|V_{cb}| & (42.0\pm 0.7)\cdot 10^{-3} & \cite{Buchmuller} &
\mbox{inclusive $b\to c \ell\bar\nu_\ell$}\\
|V_{ub}|_{\rm incl} & (4.4\pm 0.3)\cdot 10^{-3} & \cite{HFAG} &
\mbox{our average}\\
|V_{ub}|_{\rm excl} & (3.8\pm 0.6)\cdot 10^{-3} & \cite{HFAG} &
\mbox{our average}\\
\gamma & (65\pm 20)^\circ & \cite{UTfit} & \mbox{UTfit
  average}\\\hline
R_b^{\rm incl} & 0.45\pm 0.03 &  \mbox{Eq.~(\ref{Rb-def})} & \\
R_b^{\rm excl} & 0.39\pm 0.06 &  \mbox{Eq.~(\ref{Rb-def})}& \\
R_t& 0.91 \pm 0.16& \mbox{Eq.~(\ref{Rt})} & \mbox{error dominated by $\gamma$}\\
|V_{td}^* V_{tb}|_{\rm incl} & (8.6\pm 1.5)\cdot 10^{-3} &   \mbox{Eq.~(\ref{CKM-Bd})} 
& \mbox{error dominated by $\gamma$}\\
|V_{td}^* V_{tb}|_{\rm excl} & (8.6\pm 1.3)\cdot 10^{-3} &   \mbox{Eq.~(\ref{CKM-Bd})} 
& \mbox{error dominated by $\gamma$}\\
|V_{ts}^* V_{tb}| & (41.3\pm 0.7)\cdot 10^{-3} &
\mbox{Eq.~(\ref{Vts})} & \\
\beta_{\rm incl} & (26.7\pm 1.9)^\circ & \mbox{Eq.~(\ref{beta-true})}
& \mbox{error dominated by $R_b$}\\
\beta_{\rm excl} & (22.9\pm 3.8)^\circ & \mbox{Eq.~(\ref{beta-true})}
& \mbox{error dominated by $R_b$}\\
\hline
\end{array}
$$
\caption[]{CKM parameters used in our analysis. All parameters are
  determined using input from tree-level processes only and the
  unitarity of the CKM matrix.}\label{tab:CKM}
\bigskip
$$
\begin{array}{|l||c|}
\hline
{\rm Parameter} & {\rm Value} \\\hline
\lambda & 0.225\pm0.001\\
|V_{cb}| & (42.0\pm 0.7)\cdot 10^{-3}\\
|V_{ub}| & (4.4\pm 0.2)\cdot 10^{-3}\\
\gamma & (70\pm 5)^\circ\\\hline
R_b & 0.45\pm 0.02\\
R_t & 0.95\pm 0.04\\
|V_{td}^* V_{tb}| & (8.9\pm 0.4)\cdot 10^{-3}\\
|V_{ts}^* V_{tb}| & (41.3\pm 0.7)\cdot 10^{-3}\\
\beta & (26.6\pm 1.2)^\circ\\\hline
f_{B_d} \hat B_{B_d}^{1/2} &(0.244\pm 0.026)\,{\rm GeV}\\
f_{B_s} \hat B_{B_s}^{1/2} & (0.295\pm 0.036)\,{\rm GeV}\\
\rho_d & 0.69\pm 0.16\\
\rho_s & 0.74\pm 0.18\\
\xi &  1.210^{+0.047}_{-0.035}\\
\rho_s/\rho_d & 1.07\pm0.12 \\\hline
\end{array}
$$
\renewcommand{\arraystretch}{1}
\addtolength{\arraycolsep}{-3pt}
\caption[]{Benchmark values and uncertainties for CKM and hadronic
  parameters in 2010.}\label{tab:2010}
\end{table}

\begin{figure}[t]
$$\epsfxsize=0.47\textwidth\epsffile{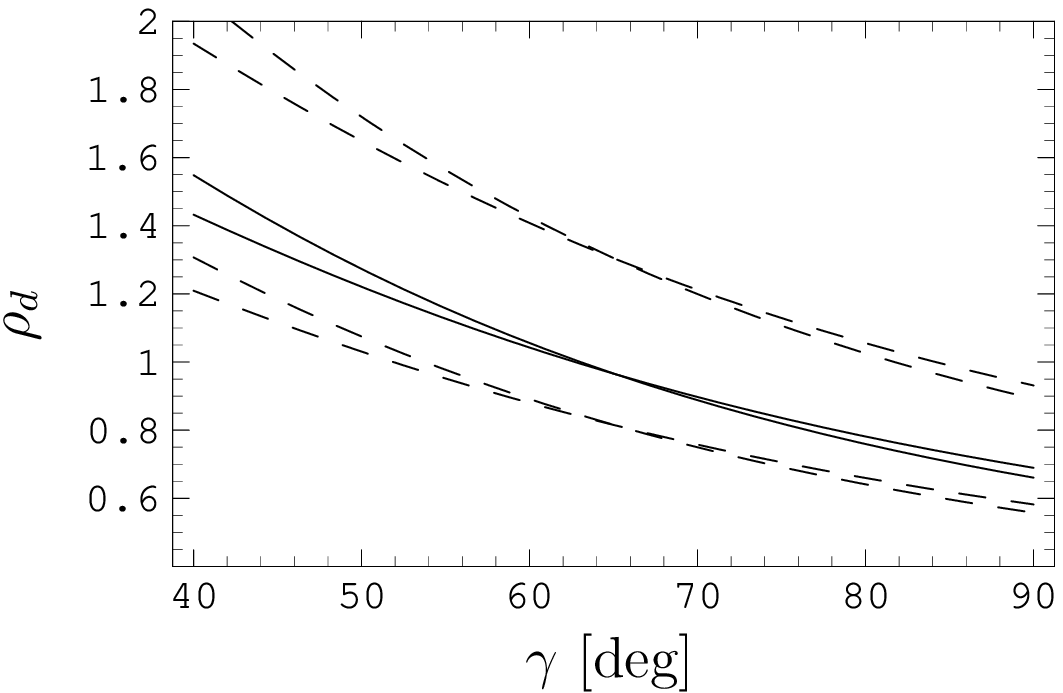}\quad
\epsfxsize=0.47\textwidth\epsffile{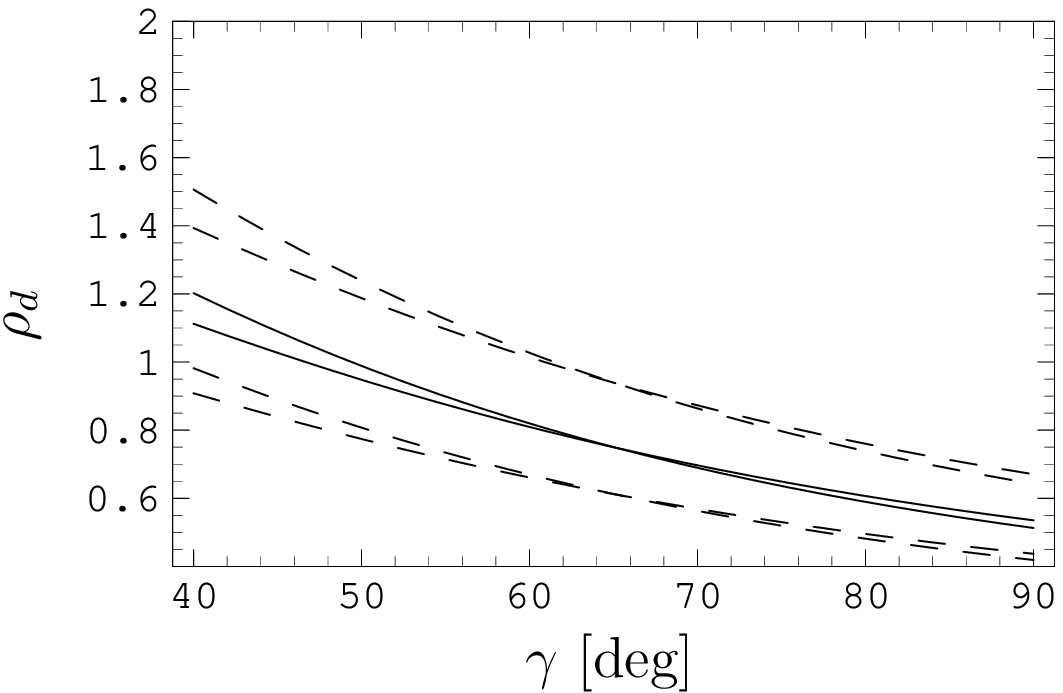}
$$
 \vspace*{-1truecm}
\caption[]{The dependence of $\rho_d$ on $\gamma$ for $R_b=(0.39,0.45)$
  and various values of $f_{B_d}\hat{B}_{B_d}^{1/2}$. Left panel: JLQCD results
  (\ref{JLQCD}):
$f_{B_d}\hat{B}_{B_d}^{1/2}=0.215\,$GeV (solid
lines), $f_{B_d}\hat{B}_{B_d}^{1/2}=(0.185,0.234)\,$GeV (dashed
lines). Right panel: ditto for (HP+JL)QCD results (\ref{HPQCD}):  
$f_{B_d}\hat{B}_{B_d}^{1/2}=0.244\,$GeV (solid
lines), $f_{B_d}\hat{B}_{B_d}^{1/2}=(0.218,0.270)\,$GeV (dashed
lines).}\label{fig:DMd}
\end{figure}

In Fig.~\ref{fig:DMd}, we illustrate the dependence of $\rho_d$ defined in
(\ref{rhod-def}) on $\gamma$, $R_b$ and $f_{B_d}\hat{B}_{B_d}^{1/2}$. It is 
evident that $\rho_d$ depends rather strongly on $\gamma$ and
$f_{B_d}\hat{B}_{B_d}^{1/2}$, but less so on $R_b$. For the two different
lattice results, we  obtain
\begin{eqnarray}
\left.\Delta M_d^{\rm SM}\right|_{\rm JLQCD} & = & \left[0.52\pm
0.17(\gamma,R_b)^{-0.09}_{+0.13}(f_{B_d}
  \hat B_{B_d}^{1/2})\right]\,{\rm ps}^{-1}\,,\nonumber\\
\left.\rho_d\right|_{\rm JLQCD} &=&
  0.97\pm0.33(\gamma,R_b)^{-0.17}_{+0.26}(f_{B_d}
  \hat B_{B_d}^{1/2})\,,\nonumber\\
\left.\Delta M_d^{\rm SM}\right|_{\rm (HP+JL)QCD} & = & \left[0.69\pm
0.13(\gamma,R_b)\pm 0.08(f_{B_d}
  \hat B_{B_d}^{1/2})\right]\,{\rm ps}^{-1}\,,\nonumber\\
\left.\rho_d\right|_{\rm  (HP+JL)QCD} &=& 
0.75\pm0.25(\gamma,R_b)\pm0.16(f_{B_d}  
\hat B_{B_d}^{1/2})\,,\label{DMd-SM-numbers}
\end{eqnarray}
where we made explicit the errors arising from the uncertainties of $(\gamma$, $R_b)$ and 
$f_{B_d}\hat B_{B_d}^{1/2}$. These results are compatible with the 
SM value $\rho_d=1$, but suffer from considerable uncertainties, which presently 
leave sizeable room for NP contributions to $\Delta M_d$; we
shall quantify below the allowed values of $\kappa_d$ and $\sigma_d$ following from
the contours in Fig.~\ref{fig:kappa-rho-d}.

\begin{figure}[t]
$$\epsfxsize=0.47\textwidth\epsffile{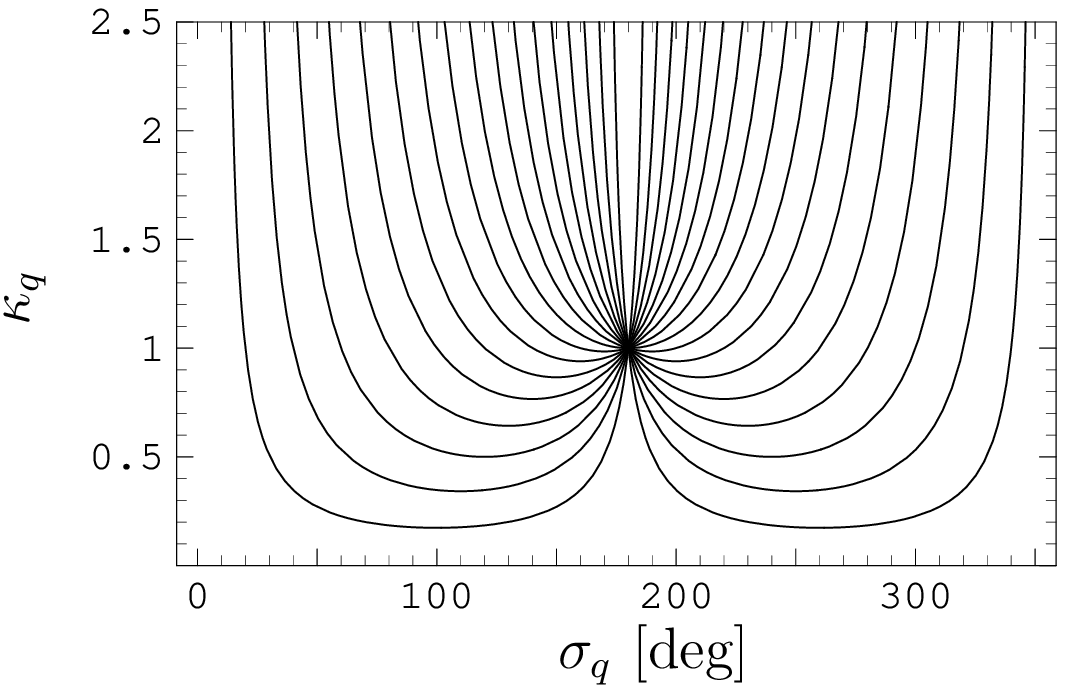}$$
 \vspace*{-1truecm}
\caption[]{The dependence of $\kappa_q$ on $\sigma_q$ for values of 
$\phi_q^{\rm NP}$ varied between $\pm10^\circ$ (lower curves) and 
$\pm170^\circ$ in steps of $10^\circ$: the curves for $0^\circ<\sigma_q<180^\circ$
and $180^\circ<\sigma_q<360^\circ$ correspond to positive and negative values
of $\phi_q^{\rm NP}$, respectively. The contours apply  to both the $q=d$ and the 
$q=s$ system.
 }\label{fig:kappa-sigma}
$$\epsfxsize=0.47\textwidth\epsffile{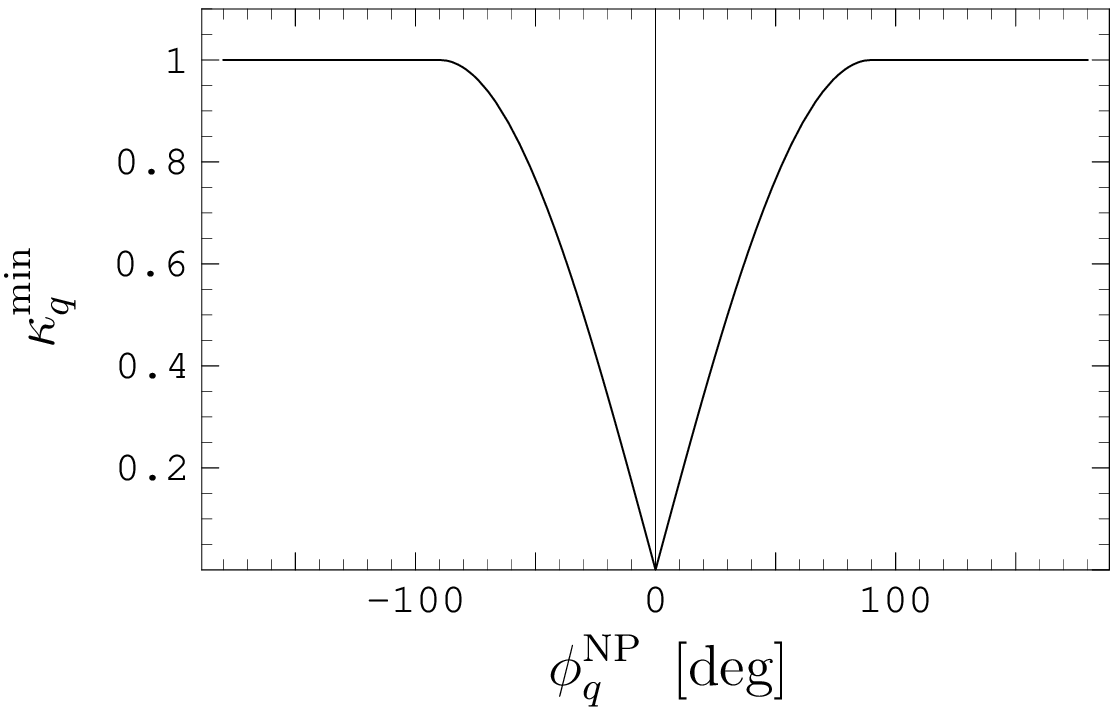}$$
\vspace*{-1cm}
\caption[]{The minimum value $\kappa_q^{\rm min}$ of $\kappa_q$ as
  function of the NP mixing phase $\phi_q^{\rm NP}$.}\label{fig:kappamin}
\end{figure}

\boldmath
\subsection{Constraints on NP through CP Violation: $\phi_d$}\label{ssec:phi-d}
\unboldmath
The second constraint on the allowed values of $\kappa_d$ and
$\sigma_d$ is provided by the experimental value of the $B_d$ mixing phase
$\phi_d= \phi_d^{\rm SM} + \phi_d^{\rm NP}$. 
Using (\ref{phi-d}), a given value of $\phi_d^{\rm NP}$ allows one to determine 
$\kappa_d$ as a function of $\sigma_d$ with the help of the following
expressions, which hold again in the general case $q\in\{d,s\}$:
\begin{equation}
\kappa_q=\frac{\tan\phi_q^{\rm NP}}{\sin\sigma_q-\cos\sigma_q\tan\phi_q^{\rm NP}}\,,
\end{equation}
\begin{equation}
\sin\phi_q^{\rm NP}=\frac{\kappa_q\sin\sigma_q}{\sqrt{1+2\kappa_q\cos\sigma_q
+\kappa_q^2}}\,, 
\quad
\cos\phi_q^{\rm NP}=\frac{1+\kappa_q\cos\sigma_q}{\sqrt{1+2\kappa_q\cos\sigma_q
+\kappa_q^2}}\,.
\end{equation}
In Fig.~\ref{fig:kappa-sigma}, we  illustrate the corresponding contours for 
various values of $\phi_q^{\rm NP}$. Note in particular that $\kappa_q$ 
is bounded from below for any given value of $\phi_q^{\rm NP}\not=0$. 
The relation between the allowed values of $\phi^{\rm NP}_q$ and $\kappa_q$ 
is given by
\begin{equation}\label{phimaxmin}
\phi_q^{\rm NP,max(min)} = \arg \left\{ 1 + \kappa_q \left(-\kappa_q^2 \pm
  i \sqrt{1-\kappa_q^2}\right)\right\},
\end{equation}
i.e.\ for any  non-zero value of $\phi_q^{\rm NP}$, $\kappa_q$ must be larger 
than the minimum value plotted in Fig.~\ref{fig:kappamin}.

In order to make use of these theoretically clean contours, one needs  to determine
the NP phase $\phi_q^{\rm NP}$. As is well known, $\phi_d$ can be experimentally 
accessed in the mixing-induced CP asymmetry of the ``golden" decay 
$B^0_d\to J/\psi K_{\rm S}$ (and similar $b\to c\bar c s$ charmonium modes)
\cite{bisa}. The most recent average of the $B$-factory data for such transitions
obtained by  HFAG is \cite{HFAG}
\begin{equation}\label{s2b-exp}
(\sin\phi_d)_{c\bar cs}=0.687\pm0.032\,.
\end{equation}
In principle, this quantity could be affected by NP contributions to both
$B^0_d$--$\bar B^0_d$ mixing and  $b\to c\bar cs$ 
decay amplitudes \cite{FM-BpsiK,RF-land}. A probe of the latter effects is provided by 
decays like $B_d\to D\pi^0, D\rho^0,\dots$, which are pure tree decays 
and do not receive any penguin contributions. If the neutral $D$ mesons
are observed through their decays into CP eigenstates $D_\pm$, these decays
allow an extremely clean determination of the ``true" value of $\sin\phi_d$ 
\cite{RF-BdDpi0}.  A possible discrepancy with $(\sin\phi_d)_{c\bar cs}$
would be attributed to NP contributions to the $b\to c\bar cs$ decay amplitudes.
Consequently, detailed feasibility studies for the exploration of 
$B_d\to D\pi^0, D\rho^0, ...$\ modes at a super-$B$ factory are strongly 
encouraged. In this paper, however, we assume that NP effects entering decay
amplitudes are negligible. Eq.~(\ref{phi-d}) then gives the following
expression:
\begin{equation}
(\sin\phi_d)_{c\bar cs}\equiv\sin\phi_d=\sin(2\beta+\phi_d^{\rm NP})\,.
\end{equation}
The experimental value (\ref{s2b-exp}) yields the twofold solution
\begin{equation}\label{phid-det}
\phi_d=(43.4\pm2.5)^\circ \quad\lor\quad (136.6\pm2.5)^\circ,
\end{equation}
where the latter result is in dramatic conflict with global CKM fits and
would require a large NP contribution to $B^0_d$--$\bar B^0_d$ mixing 
\cite{FlMa,FIM}. However, experimental information on the sign of $\cos\phi_d$ 
rules out a negative value of this quantity at greater than 95\% C.L.\
\cite{WG5-report}, so that we are left with $\phi_d=(43.4\pm2.5)^\circ$.

\begin{figure}[t]
$$
\epsfxsize=0.47\textwidth\epsffile{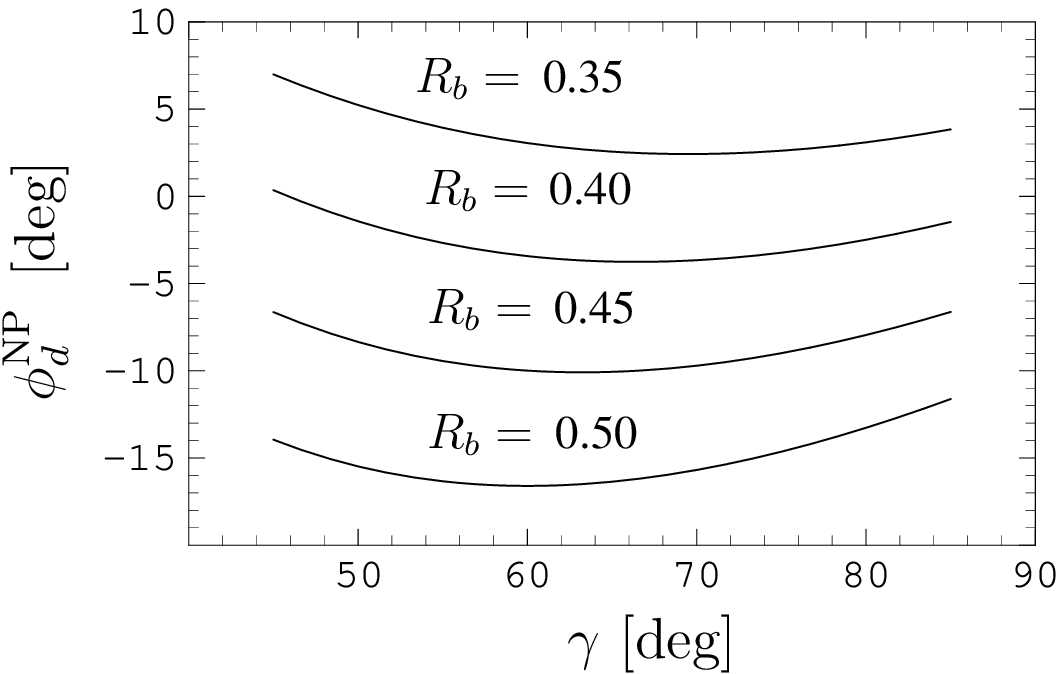}\qquad
\epsfxsize=0.47\textwidth\epsffile{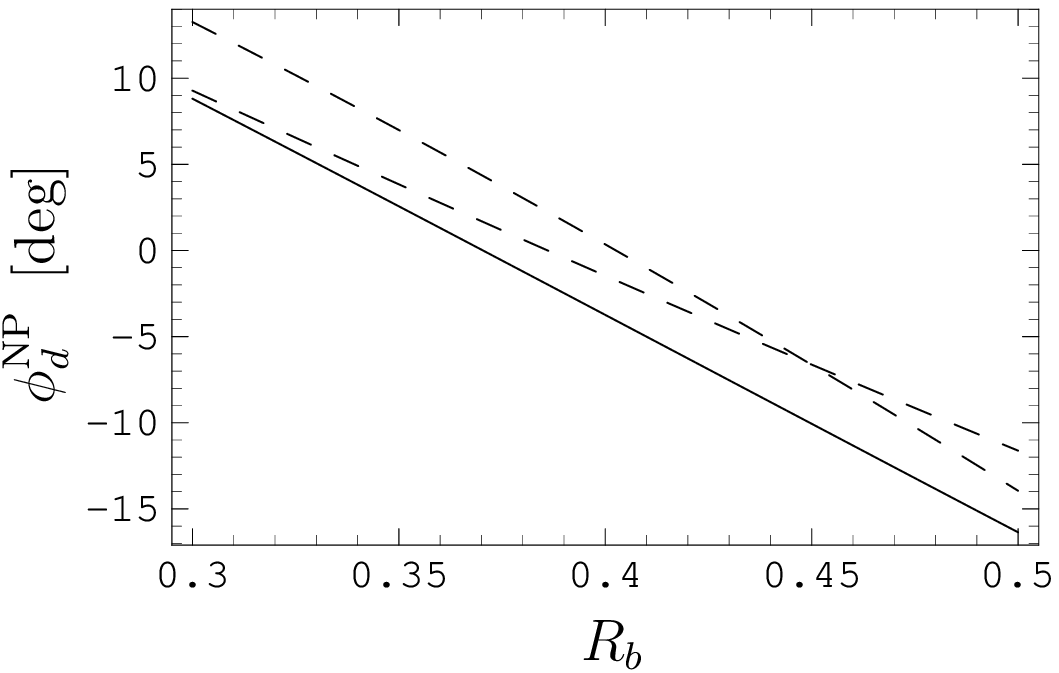}
$$
 \vspace*{-1truecm}
\caption[]{The determination of  $\phi_d^{\rm NP}$ for 
$\phi_d=43.4^\circ$.  Left panel: $\phi_d^{\rm NP}$ as a function of 
$\gamma$ for various values of $R_b$. Right panel: $\phi_d^{\rm NP}$ 
as a function of $R_b$ for various values of $\gamma$ (solid line:
$\gamma=65^\circ$, dashed lines: $\gamma=(45^\circ,85^\circ)$).
}\label{fig:phi-NP-d}
\end{figure}

The SM prediction of the mixing phase, $\phi_d^{\rm SM}=2\beta$,
 Eq.~(\ref{phi-SM}), can easily be obtained in terms of the tree-level quantities 
$R_b$ and $\gamma$, as
\begin{equation}\label{beta-true}
\sin\beta=\frac{R_b\sin\gamma}{\sqrt{1-2R_b\cos\gamma+R_b^2}}\,, \quad
\cos\beta=\frac{1-R_b\cos\gamma}{\sqrt{1-2R_b\cos\gamma+R_b^2}}\,.
\end{equation}
Using Eq.~(\ref{phi-d}), the experimental value of $\phi_d$ can then
immediately be converted into a result for the NP phase $\phi_d^{\rm
  NP}$, which depends on $\gamma$ and $R_b$ as illustrated in
Fig.~\ref{fig:phi-NP-d}. 
It is evident that the dependence of
$\phi^{\rm NP}_d$ on $\gamma$ is very small and that $R_b$ plays
actually the key r\^ole for its determination. Hence, we have a situation 
complementary to that shown in Fig.~\ref{fig:DMd}, where
the main dependence 
was on $\gamma$. The parameters collected in Tab.~\ref{tab:CKM} 
yield
\begin{equation}
\left.\phi_d^{\rm SM}\right|_{\rm incl} = 
(53.4\pm 3.8)^{\circ}\,,\qquad \left.\phi_d^{\rm SM}\right|_{\rm excl} =
(45.8\pm 7.6)^{\circ}\,,
\end{equation}
corresponding to 
\begin{equation}\label{phiNPd-num}
\left.\phi^{\rm NP}_d\right|_{\rm incl} = -(10.1\pm 4.6)^\circ\,,\qquad
\left.\phi^{\rm NP}_d\right|_{\rm excl} = -(2.5\pm 8.0)^\circ\,;
\end{equation}
results of $\phi_d^{\rm NP}\approx-10^\circ$ were also recently obtained in 
Refs.~\cite{BFRS-05,UTfit-NP}. Note that the emergence of a  non-zero value of
$\phi_d^{\rm NP}$ is caused by the large value of $|V_{ub}|$ from
inclusive semileptonic decays, but that $\phi_d^{\rm NP}$ is
compatible with zero for $|V_{ub}|$ from exclusive decays.
The consequences of the presence of a small NP phase $\phi_d^{\rm NP}\approx-10^\circ$
are rather dramatic: from Fig.~\ref{fig:kappamin}, one reads off the sizeable lower 
bound $\kappa_d\gsim 0.17$. Although this result hinges on the value of
$|V_{ub}|_{\rm incl}$, and hence presently is not conclusive, the underlying 
reasoning also applies to the $B_s$ system: even a small NP phase 
$\phi_s^{\rm NP}$ implies considerable NP contributions to the mixing matrix 
element $M^s_{12}$.

\begin{figure}[t]
$$\epsfxsize=0.47\textwidth\epsffile{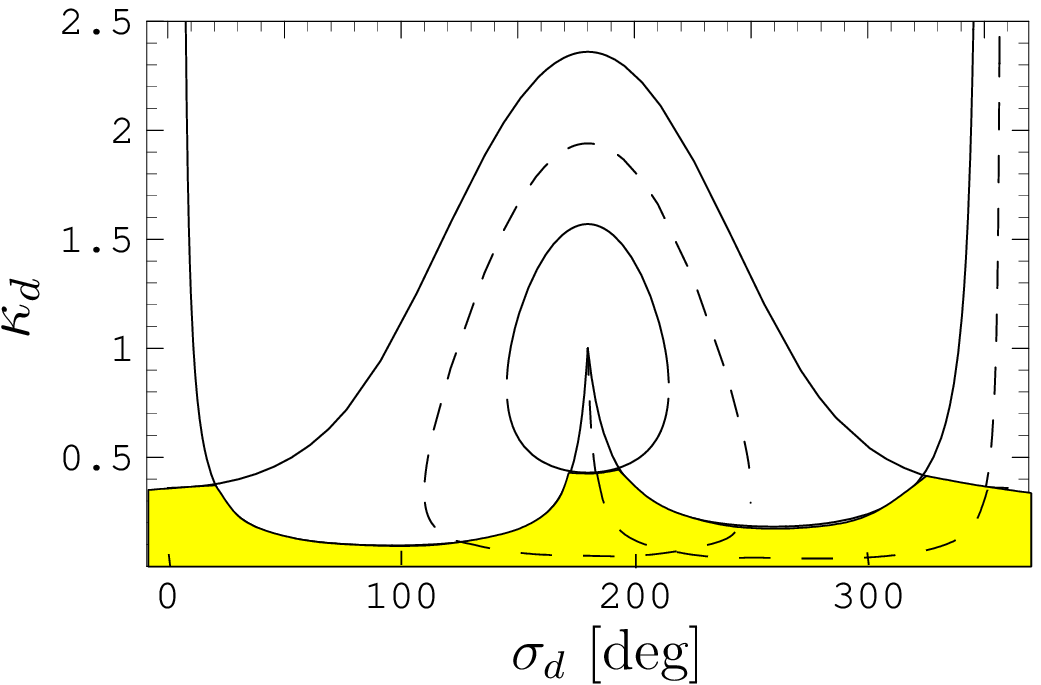}\quad
\epsfxsize=0.47\textwidth\epsffile{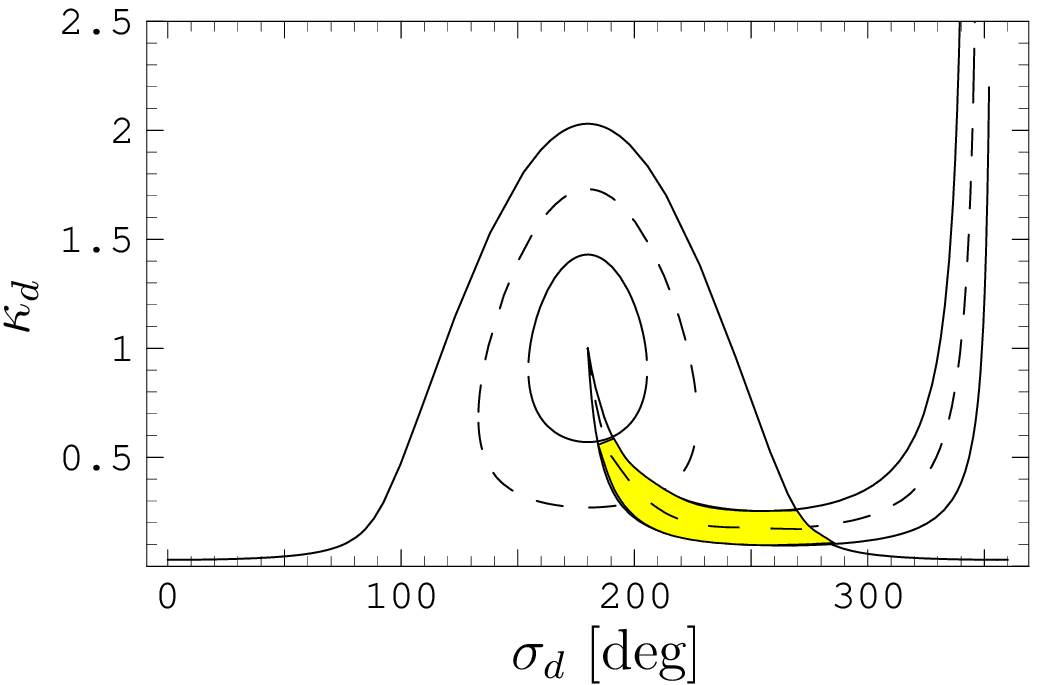}
$$
 \vspace*{-1truecm}
\caption[]{Left panel: allowed region (yellow/grey) in the $\sigma_d$--$\kappa_d$
  plane in a scenario with the JLQCD lattice results (\ref{JLQCD}) and 
  $\left.\phi^{\rm NP}_d\right|_{\rm excl}$. Dashed lines: central values of $\rho_d$ 
  and $\phi^{\rm NP}_d$, solid lines: $\pm 1\,\sigma$. Right panel: ditto for the 
 scenario with the (HP+JL)QCD   lattice results
  (\ref{HPQCD}) and  $\left.\phi^{\rm NP}_d\right|_{\rm incl}$. 
}\label{fig:res-k-sig-d}
$$\epsfxsize=0.47\textwidth\epsffile{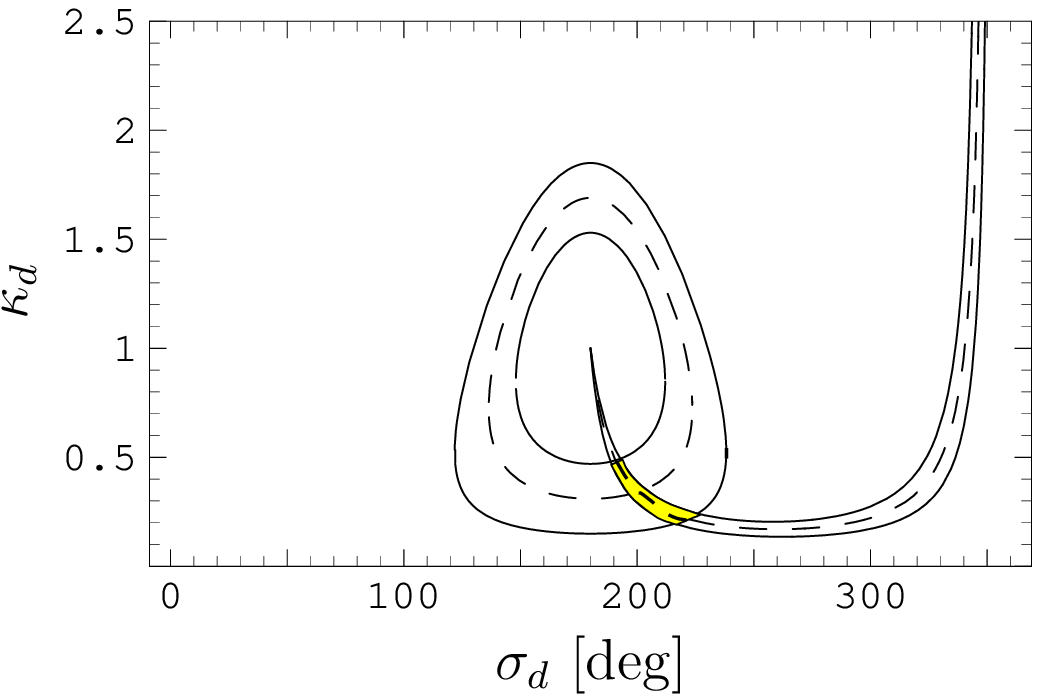}$$
\vspace*{-1cm}
\caption[]{Allowed region in the $\sigma_d$--$\kappa_d$ plane
  (yellow/grey) in our 2010 scenario, using the parameters collected in 
Tab.~\ref{tab:2010} and $\phi^{\rm NP}_d = -(9.8\pm 2.0)^\circ$.}\label{fig:future-d}
\end{figure}

\boldmath
\subsection{Combined Constraints on NP through $\Delta M_d$ and
  $\phi_d$: 2006 and 2010}\label{ssec:comb-Bd}
\unboldmath
We are now finally in a position to combine the constraints from both
$\Delta M_d$ and $\phi_d$ to constrain the allowed region in the 
$\sigma_d$--$\kappa_d$ plane. The corresponding results are shown
in Fig.~\ref{fig:res-k-sig-d}, demonstrating the power of the contours
described in the previous subsections for a transparent determination of
$\sigma_d$ and $\kappa_d$. We see that a non-vanishing value
of $\phi_d^{\rm NP}$, even as small as $\phi_d^{\rm NP}\approx -10^\circ$,
has a strong impact on the allowed space in the $\sigma_d$--$\kappa_d$ plane.
In both scenarios with different lattice results and different
values for $|V_{ub}|$, the upper bounds of $\kappa_d\lsim2.5$ on the NP
contributions following from the experimental value of $\Delta M_d$ are reduced
to $\kappa_d\lsim0.5$. Values of this order of magnitude are expected, for instance, 
on the basis of generic field-theoretical considerations \cite{FM-BpsiK,FIM},
as well as in a recently proposed framework for ``next-to-minimal flavour
violation" \cite{NMFV,LPP}.

In order to determine $\kappa_d$ more precisely, it is mandatory
to reduce the errors of $\rho_d$,  which come from both $\gamma$ and
lattice calculations. As we noted above, the value of $\gamma$ can be
determined -- with impressive accuracy -- at the LHC \cite{schneider}, whereas
progress on the lattice side is much harder to predict, but will hopefully be made. 
Assuming our benchmark scenario of Tab.~\ref{tab:2010}, which corresponds to 
the lattice results of Eq.~(\ref{HPQCD}), the $\sigma_d$--$\kappa_d$ plane in 
2010 looks like shown in Fig.~\ref{fig:future-d} -- and actually implies $5\,\sigma$ 
evidence for NP from 
$\phi_d^{\rm NP} = -(9.8\pm 2.0)^\circ$. 
Although there is only a small allowed region left, $\kappa_d$ is still only badly 
constrained; for
an extraction with  10\% uncertainty, $f_{B_d}\hat B_{B_d}^{1/2}$ is required
to 5\% accuracy, i.e.\  the corresponding error in (\ref{HPQCD}) has to be reduced 
by a factor of 2, which is the benchmark lattice theorists should strive for.

\boldmath
\section{The $B_s$-Meson System}\label{sec:Bs}
\unboldmath
\boldmath
\subsection{Constraints on NP through $\Delta M_s$}
\unboldmath
Let us now have a closer look at the $B_s$-meson system. In order to describe
NP effects in a model-independent way, we parametrize them analogously
to (\ref{M-d}) and (\ref{phi-d}). The relevant CKM factor is
$|V_{ts}^* V_{tb}|$. Using once again the
unitarity of the CKM matrix and including next-to-leading
order terms in the Wolfenstein expansion as given in Ref.~\cite{blo}, we have
\begin{equation}\label{Vts}
\left|\frac{V_{ts}}{V_{cb}}\right|=1-\frac{1}{2}\left(1-2R_b\cos\gamma\right)\lambda^2
+{\cal O}(\lambda^4).
\end{equation}
Consequently, apart from the tiny correction in $\lambda^2$, the  CKM
factor for $\Delta M_s$ is independent of $\gamma$ and $R_b$,
which is an important advantage in comparison with the $B_d$-meson system.
The accuracy of the SM prediction of $\Delta M_s$ is hence  limited by the
hadronic mixing parameter $f_{B_s}\hat{B}_{B_s}^{1/2}$. Using the numerical
values discussed in Section~\ref{sec:input}, we obtain
\begin{eqnarray}
\left.\Delta M_s^{\rm SM}\right|_{\rm JLQCD} & = & (16.1\pm 2.8) \,{\rm ps}^{-1}\,,\nonumber\\
\left.\rho_s\right|_{\rm JLQCD} &=&
1.08^{+0.03}_{-0.01} \mbox{(exp)} \pm 0.19 \mbox{(th)}\,,\nonumber\\
\left.\Delta M_s^{\rm SM}\right|_{\rm (HP+JL)QCD} & = & (23.4\pm 
3.8)\,{\rm ps}^{-1}\,,\nonumber\\
\left.\rho_s\right|_{\rm  (HP+JL)QCD} &=& 
0.74^{+0.02}_{-0.01} \mbox{(exp)} \pm 0.18 \mbox{(th)}\,,\label{36}
\end{eqnarray}
where we made the experimental and theoretical errors explicit. The
values of $\rho_s$, which is  defined in analogy to  (\ref{rhod-def}), 
refer to the CDF measurement of $\Delta M_s$ in (\ref{exp}).
These numbers are consistent with the
SM case $\rho_s=1$, but suffer from significant theoretical uncertainties, 
which are much larger than the experimental errors. Nevertheless, it is
interesting to note that the (HP+JL)QCD result is $1.5\,\sigma$ below the SM;
a similar pattern arises in (\ref{DMd-SM-numbers}), though at the $1\,\sigma$ level.
Any more precise statement about the presence or absence of NP
requires the reduction of theoretical uncertainties. 

\begin{figure}[t]
$$\epsfxsize=0.47\textwidth\epsffile{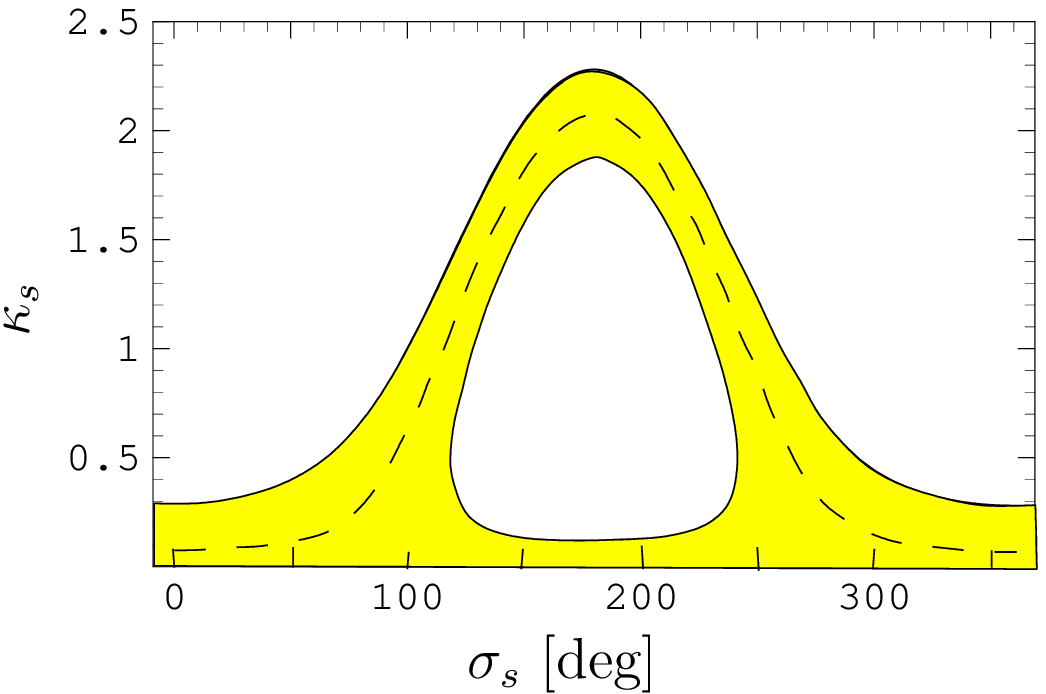}\quad
\epsfxsize=0.47\textwidth\epsffile{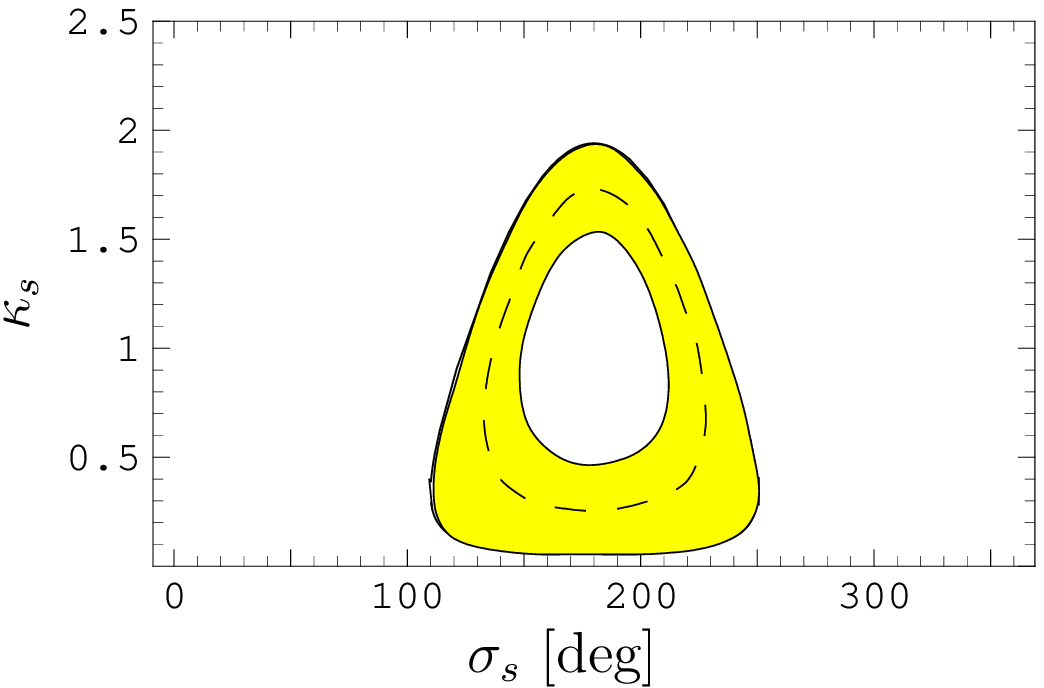}
$$
 \vspace*{-1truecm}
\caption[]{The allowed regions (yellow/grey) in the $\sigma_s$--$\kappa_s$ plane.
Left panel: JLQCD lattice results (\ref{JLQCD}). Right panel: (HP+JL)QCD lattice 
results (\ref{HPQCD}).}\label{fig:MDs-NP}
\end{figure}

In Fig.~\ref{fig:MDs-NP}, we show the constraints in the $\sigma_s$--$\kappa_s$ 
plane, which can be obtained from $\rho_s$ with the help of the $B_s$ counterpart 
of (\ref{kappa-sig-rho}). We see that  upper bounds of $\kappa_s\lsim 2.5$
arise from the measurement of $\Delta M_s$. In the case of (\ref{36}), the 
bound on $\sigma_s$ following from (\ref{sigma-bound}) would interestingly be 
effective, and imply $110^\circ\leq\sigma_s\leq250^\circ$. Consequently, 
the CDF measurement of $\Delta M_s$ leaves ample space for the 
NP parameters $\sigma_s$ and $\kappa_s$. This situation will change significantly 
as soon as information about CP violation in the $B_s$-meson system becomes
available. We shall return to this topic in Subsection~\ref{ssec:Bs-CP}.

\boldmath
\subsection{Constraints on NP through $\Delta M_s$ and $\Delta M_d$}
\unboldmath
It is interesting to consider the ratio of $\Delta M_s$ and $\Delta M_d$,
which can be written as follows:
\begin{equation}\label{DMs-DMd-rat}
\frac{\Delta M_s}{\Delta M_d} =  \frac{\rho_s}{\rho_d}
\left|\frac{V_{ts}}{V_{td}}\right|^2 \frac{M_{B_s}}{M_{B_d}}\, \xi^2\,,
\end{equation}
where the hadronic $SU(3)$-breaking parameter $\xi$ is defined
in Subsection~\ref{ssec:hadr-uncert}. In the class of NP models with
``minimal flavour violation" \cite{MFV},\footnote{See Ref.~\cite{buras-MFV} for a 
review, and Ref.~\cite{BBGT} for a recent analysis addressing also the
$\Delta M_s$ measurement.} 
which contains also the SM, we have $\rho_s/\rho_d=1$, so that (\ref{DMs-DMd-rat}) 
allows the extraction of the CKM factor $|V_{ts}/V_{td}|$, and hence
$|V_{td}|$, as $|V_{ts}|$ is known -- to excellent accuracy -- from (\ref{Vts}). The advantage of this
determination lies in the reduced theoretical uncertainty of $\xi$ as compared to 
$f_{B_d}\hat B_{B_d}^{1/2}$.

In this paper, however, we turn the tables and constrain
the ratio $\rho_s/\rho_d$ through $\Delta M_s/\Delta M_d$. To this end, we
express -- in analogy to (\ref{CKM-Bd})  -- the UT side 
\begin{equation}\label{Rt-def}
R_t \equiv \frac{1}{\lambda}\left|\frac{V_{td}}{V_{cb}}\right| 
=  \frac{1}{\lambda}\left|\frac{V_{td}}{V_{ts}}\right| \left[ 
1-\frac{1}{2}\left(1-2R_b\cos\gamma\right)\lambda^2+{\cal O}(\lambda^4)\right]
\end{equation}
in terms of $R_b$ and $\gamma$:
\begin{equation}\label{Rt}
R_t = \sqrt{1- 2 R_b\cos\gamma + R_b^2}\,,
\end{equation}
allowing the determination of $R_t$ through processes that are essentially
unaffected by NP. The resulting value of $R_t$ depends rather strongly on 
$\gamma$, which is the main source of uncertainty. Another
determination of $R_t$ that is independent of $\gamma$ and $R_b$ 
can, in principle, be obtained from radiative decays, in particular the ratio of 
branching ratios ${\cal B}(B\to (\rho,\omega)\gamma)/{\cal B}(B\to K^*\gamma)$, 
but is presently limited by experimental statistics; see Ref.~\cite{VtdVts} for a 
recent analysis.

\begin{figure}[t] 
$$\epsfxsize=0.47\textwidth\epsffile{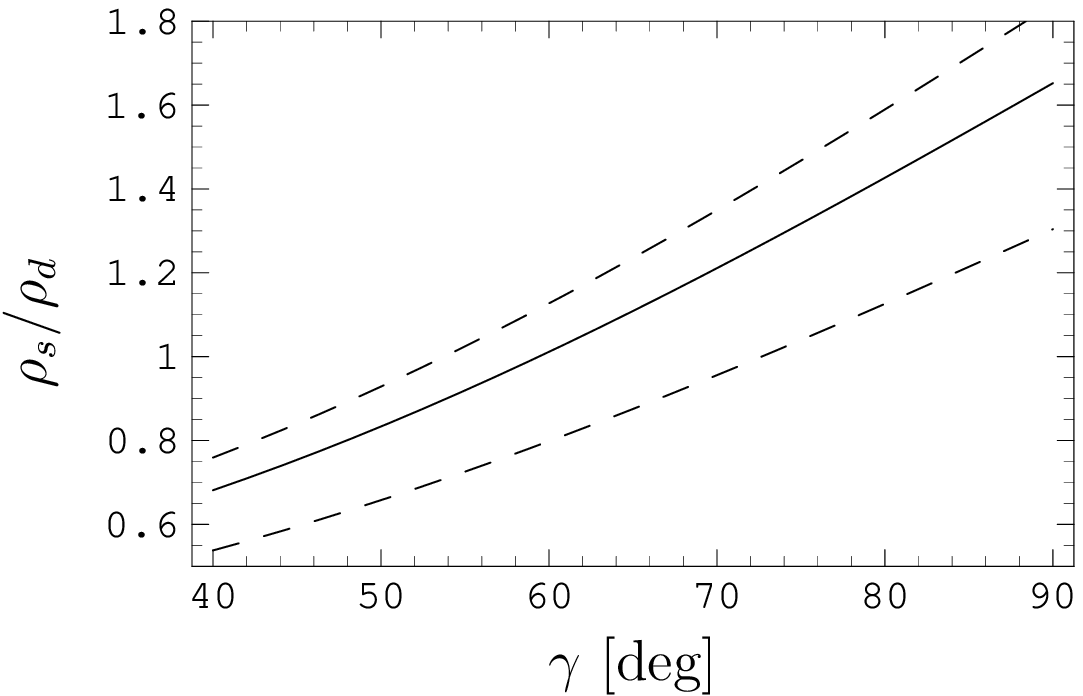}\quad
\epsfxsize=0.47\textwidth\epsffile{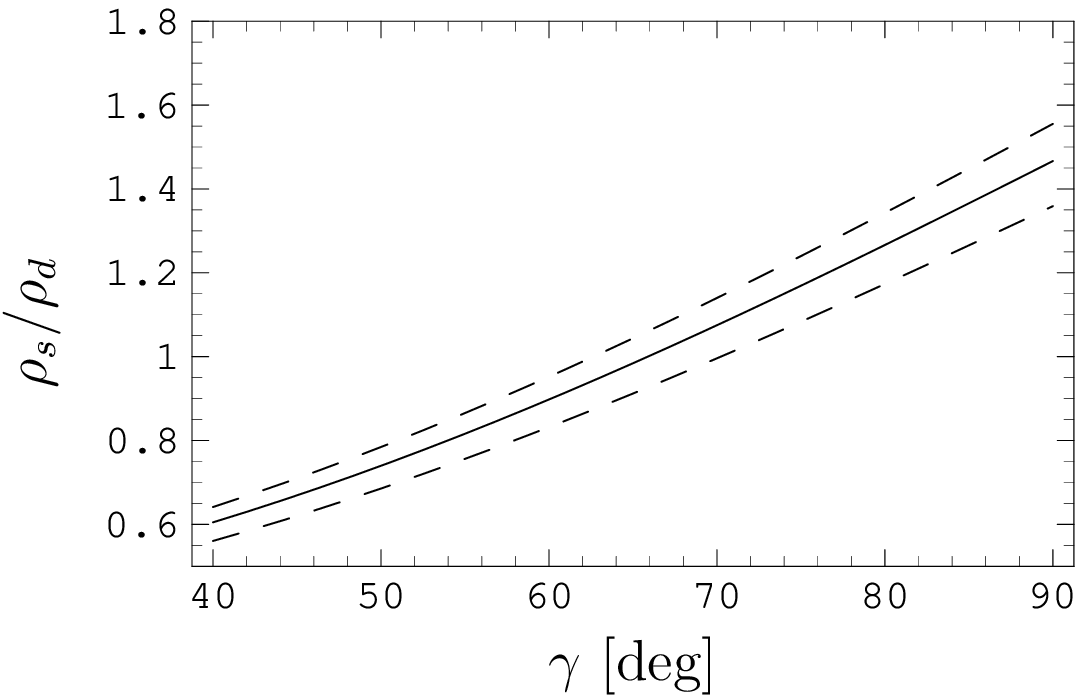}$$
\vspace*{-1cm}
   \caption[]{The dependence of $\rho_s/\rho_d$ on $\gamma$ for the central 
   values of $\Delta M_{d,s}$ in (\ref{exp}). 
Left panel: JLQCD results (\ref{JLQCD}). 
   Right panel:  (HP+JL)QCD results (\ref{HPQCD}). The plots are nearly
   independent of $R_b$.}
   \label{fig:rhos-rhod}
\end{figure}

Combining (\ref{DMs-DMd-rat}) and (\ref{Rt-def}), we obtain the
following expression for $\rho_s/\rho_d$:
\begin{equation}
\frac{\rho_s}{\rho_d}=\lambda^2\left[1-2R_b\cos\gamma+R_b^2\right]
\left[1+(1-2R_b\cos\gamma)\lambda^2+{\cal O}(\lambda^4)\right]
\frac{1}{\xi^2}\frac{M_{B_d}}{M_{B_s}}
\frac{\Delta M_s}{\Delta M_d}\,.
\end{equation}
In Fig.~\ref{fig:rhos-rhod}, we plot this ratio for the central values
of $\Delta M_d$ and $\Delta M_s$ in (\ref{exp}), as a function of the UT angle $\gamma$ for 
the values of $\xi$ given in (\ref{JLQCD}) and 
(\ref{HPQCD}). We find that the corresponding curves are nearly independent
of $R_b$ and that $\gamma$ is actually the key CKM parameter for the determination
of $\rho_s/\rho_d$. The corresponding numerical values are given by:
\begin{eqnarray}
\left.\frac{\rho_s}{\rho_d}\right|_{\rm JLQCD} &=&
1.11^{+0.02}_{-0.01} \mbox{(exp)} \pm 0.35 (\gamma,R_b)^{+0.12}_{-0.28}(\xi)\,,\nonumber\\
\left.\frac{\rho_s}{\rho_d}\right|_{\rm  (HP+JL)QCD} &=& 
0.99^{+0.02}_{-0.01} \mbox{(exp)} \pm 0.31 (\gamma,R_b)^{+0.06}_{-0.08}(\xi)\,.\label{2006}
\end{eqnarray}
Because of the large range of allowed values of $\gamma$,
Eq.~(\ref{gamma}), this ratio is currently not very stringently
constrained. This situation should, however,
improve significantly in the LHC era thanks to the impressive determination of $\gamma$ 
to be obtained at the LHCb experiment. For our 2010 scenario as
specified in Tab.~\ref{tab:2010},  which corresponds to the right panel of
Fig.~\ref{fig:rhos-rhod} with $\gamma=(70\pm5)^\circ$, we find:
\begin{equation}\label{rhos-rhod-2010}
\left.\frac{\rho_s}{\rho_d}\right|_{2010}=1.07 \pm 0.09
(\gamma,R_b)^{+0.06}_{-0.08}(\xi) = 1.07\pm0.12\,,
\end{equation}
where we made the errors arising from the uncertainties of $\gamma$
and $\xi$ explicit, and, in the last step, added them in quadrature. Consequently, 
the hadronic uncertainties and those induced by $\gamma$ would now be of the same 
size, which should provide additional motivation for the lattice
community to reduce the error of $\xi$ even further. Despite the
impressive reduction of uncertainty compared to the 2006 values in
(\ref{2006}), the numerical value 
in (\ref{rhos-rhod-2010}) would still not allow a stringent test of whether
$\rho_s/\rho_d$ equals one: to establish a  $3\,\sigma$ deviation from 1, central values of
$\rho_s/\rho_d=1.4$ or 0.7 would be needed. The assumed uncertainty of
$\gamma$ of $5^\circ$ could also turn out to be too pessimistic, in which
case even more progress would be needed from the lattice side to match
the experimental accuracy. 

The result in (\ref{rhos-rhod-2010}) would not necessarily suggest that there is no 
physics beyond the SM. In fact, and as can be seen from Tab.~\ref{tab:2010}, the central 
values of $\rho_d$ and $\rho_s$ would both be smaller than one, i.e.\ would both 
deviate from the SM picture, although the hadronic uncertainties would again
not allow  us to draw definite conclusions. In order to shed further light on these 
possible NP contributions, the exploration of CP-violating effects in the $B_s$-meson 
system is essential.

\boldmath
\subsection{CP Violation in the $B_s$-System}\label{ssec:Bs-CP}
\unboldmath
To date, the CP-violating phase associated with $B^0_s$--$\bar B^0_s$ mixing 
is completely unconstrained. In the SM, it is doubly Cabibbo-suppressed, and 
can be written as follows:
\begin{equation}\label{phis-SM}
\phi_s^{\rm SM}=-2\lambda^2\eta=
-2\lambda^2R_b\sin\gamma \approx -2^\circ.
\end{equation}
Here we used again (\ref{rho-bar-eta-bar}) to express the Wolfenstein
parameter $\eta$ in terms of $R_b$ and $\gamma$. Because of the
small SM phase in (\ref{phis-SM}), $B^0_s$--$\bar B^0_s$ mixing
is particularly well suited to search for NP effects, which may well
lead to a sizeable value of $\phi_s$ \cite{NiSi,BMPR}. In order to test
the SM and  probe CP-violating NP contributions to 
$B^0_s$--$\bar B^0_s$ mixing, the decay $B^0_s\to J/\psi\phi$, which
is very accessible at the LHC \cite{LHC-report},  plays a key r\^ole. 
Thanks to mixing-induced CP violation in the time-dependent angular
distribution of the $J/\psi[\to \ell^+\ell^-]\phi[\to K^+K^-]$ decay products, 
the quantity
\begin{equation}
\sin\phi_s=\sin(-2\lambda^2R_b\sin\gamma+\phi_s^{\rm NP}) 
\end{equation}
can be measured \cite{DFN,DDF}, in analogy to the determination of $\sin\phi_d$
through $B^0_d\to J/\psi K_{\rm S}$. After one year of data taking
(which corresponds to
2 $\mbox{fb}^{-1}$), LHCb expects a measurement with the statistical accuracy 
$\sigma_{\rm stat}(\sin\phi_s)\approx 0.031$; adding modes such as 
$B_s\to J/\psi \eta, J/\psi \eta'$ and $\eta_c\phi$, 
$\sigma_{\rm stat}(\sin\phi_s)\approx 0.013$ is expected after
five years \cite{schneider}. Also ATLAS and CMS will contribute to the measurement
of $\sin\phi_s$, expecting uncertainties at the 0.1 level after one year of
data taking, which corresponds to 10 $\mbox{fb}^{-1}$ \cite{smizanska, speer}.

\begin{figure}[t]
\begin{tabular}{c@{$\quad$}c}
 \includegraphics[width=0.47\textwidth]{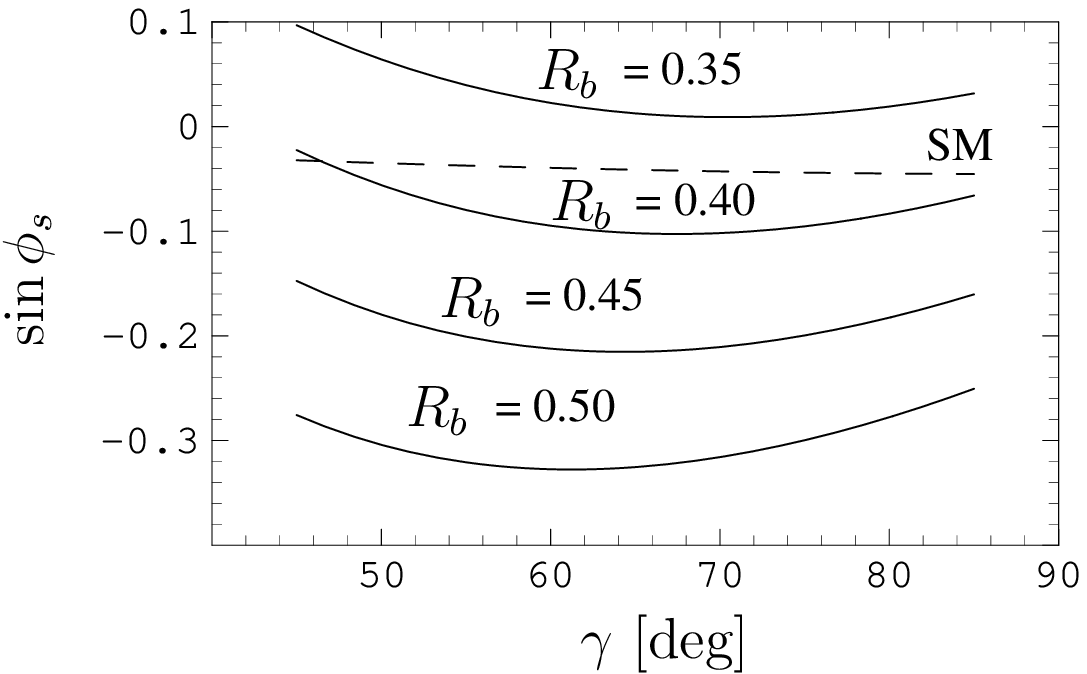} &
  \includegraphics[width=0.47\textwidth]{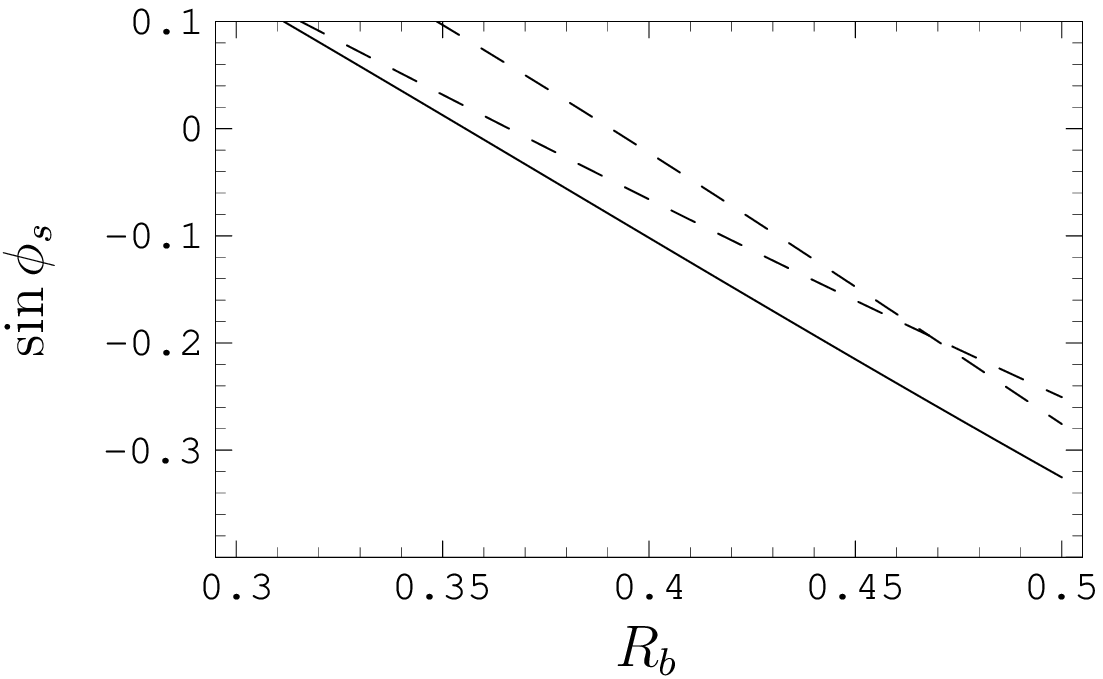}
\end{tabular}
 \vspace*{-0.3truecm}
\caption[]{$\sin\phi_s$ for a scenario with flavour-universal NP, i.e.\ $\phi_s^{\rm NP}=
  \phi_d^{\rm NP}$, as specified in Eq.~(\ref{sig-kap-rel}), and $\phi_d=43.4^\circ$.
Left panel: $\sin\phi_s$ as a function of 
$\gamma$ for various values of $R_b$. Right panel: $\sin\phi_s$ 
as a function of $R_b$ for various values of $\gamma$ (solid line:
$\gamma=65^\circ$, dashed lines: $\gamma=(45^\circ,85^\circ)$).
}\label{fig:sinPhis}
\end{figure}

In order to illustrate the impact of NP effects, let us assume that
the NP parameters satisfy the simple relation
\begin{equation}\label{sig-kap-rel}
\sigma_d=\sigma_s,  \quad \kappa_d=\kappa_s, 
\end{equation}
i.e.\ that in particular $\phi_d^{\rm NP}=\phi_s^{\rm NP}$. This scenario would
be supported by (\ref{rhos-rhod-2010}), although it would {\it not} belong to the
class of models with MFV, as new sources of CP violation would be required. 
As we have seen in the previous section, the analysis of the $B^0_d$
data for $R_b^{\rm incl}=0.45$ indicates a small NP phase around
$-10^\circ$ in the $B_d$-system. In the above scenario, that would imply the presence of
the same phase in the $B_s$-system, which would interfere constructively 
with the small SM  phase and result in  CP asymmetries at the level of $-20\%$. 
CP-violating effects of that size can easily be detected at the LHC. This 
exercise demonstrates again the great power of the $B_s$-meson system 
to reveal CP-violating NP contributions to $B^0_q$--$\bar B^0_q$ mixing.
The presence of a small NP phase could actually be considerably magnified,
as illustrated in Fig.~\ref{fig:sinPhis}. In specific NP scenarios,
also large CP-violating phases can still arise, and are in no way excluded by the 
CDF measurement of $\Delta M_s$ in (\ref{exp}).

\begin{figure}[t] 
$$\epsfxsize=0.47\textwidth\epsffile{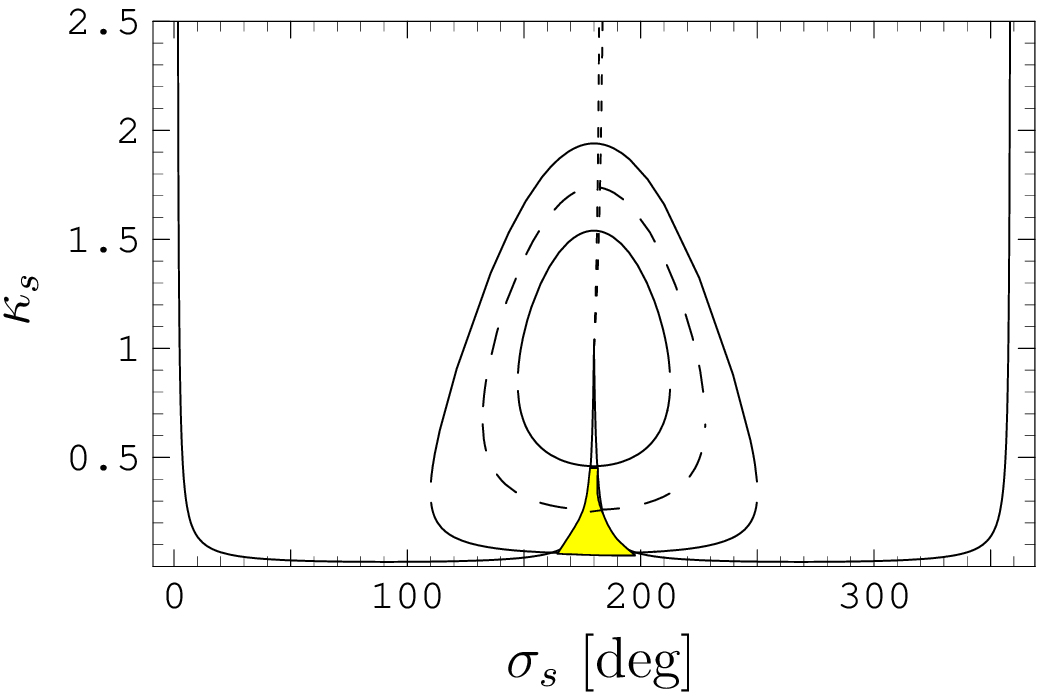}\quad
\epsfxsize=0.47\textwidth\epsffile{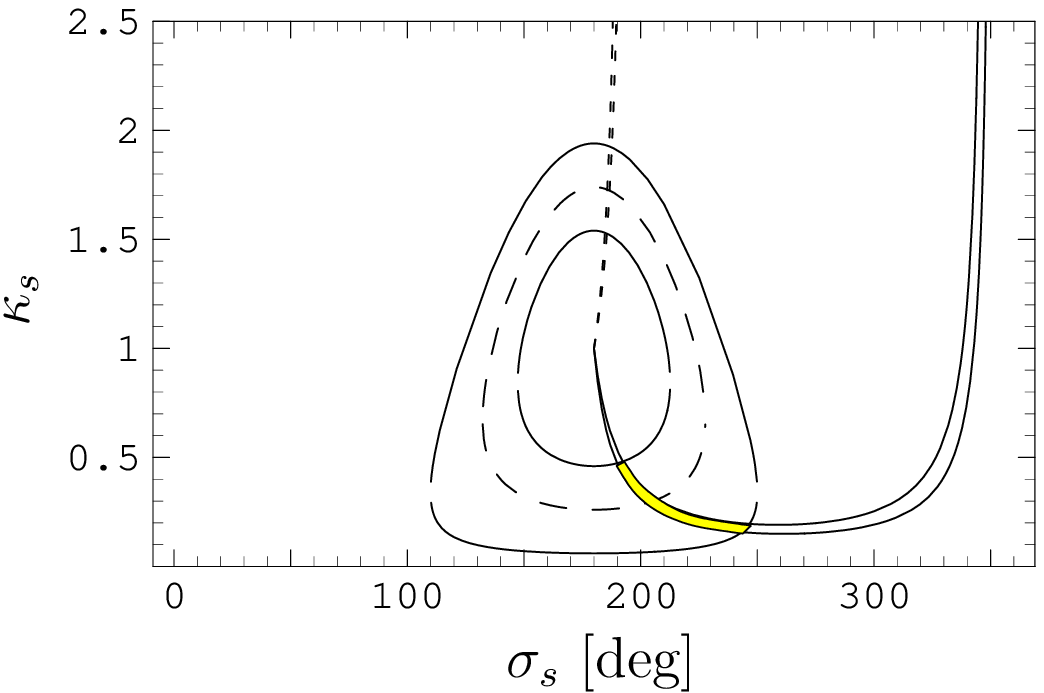}$$
\vspace*{-1cm}
   \caption[]{Combined constraints for the allowed region (yellow/grey) in the 
   $\sigma_s$--$\kappa_s$ plane through $\Delta M_s$ in (\ref{exp}) 
   for the (HP+JL)QCD results (\ref{HPQCD}) and CP violation measurements.
   Left panel: the SM scenario $(\sin\phi_s)_{\rm exp}=-0.04\pm0.02$. Right panel: 
   a NP scenario with $(\sin\phi_s)_{\rm exp}=-0.20\pm0.02$. The solid
   lines correspond to $\cos\phi_s>0$, the dotted
   lines to  $\cos\phi_s<0$.}\label{fig:sis-kas-CP}
\end{figure}

Let us finally also discuss the impact of CP violation measurements 
on the allowed region in the $\sigma_s$--$\kappa_s$ plane in our 2010
scenario. To this end, we consider two cases:
\begin{itemize}
\item[i)] $(\sin\phi_s)_{\rm exp}=-0.04\pm0.02$, in accordance with the SM;
\item[ii)] $(\sin\phi_s)_{\rm exp}=-0.20\pm0.02$, in accordance with the NP 
scenario of Fig.~\ref{fig:sinPhis}.
\end{itemize}
The measurement of $\sin\phi_s$ implies a twofold solution for $\phi_s$
and, therefore, also for $\phi_s^{\rm NP}$. However, this ambiguity can 
be resolved through the determination of the sign of $\cos\phi_s$, which
can be fixed through the strategies proposed in Ref.~\cite{DFN}. In 
Fig.~\ref{fig:sis-kas-CP}, we show the situation in the
$\sigma_s$--$\kappa_s$ plane.\footnote{The closed lines agree with those
shown in the right panel of Fig.~\ref{fig:MDs-NP}, as our 2010
scenario is based on the (HP+JL)QCD lattice results.} The dotted lines refer to negative
values of $\cos\phi_s$. Assuming that these are experimentally excluded,
we are left with strongly restricted regions, although $\kappa_s$ could still
take sizeable ranges, with upper bounds $\kappa_s\approx0.5$.
In the SM-like scenario, values of $\sigma_s$ around
$180^\circ$ would arise, i.e.\ a NP contribution with a sign opposite to 
the SM. However, due to the absence of new CP-violating effects, 
the accuracy of lattice results would have to be considerably improved
in order to allow the extraction of a value of  $\kappa_s$ incompatible with 0.
On the other hand, a measurement of $(\sin\phi_s)_{\rm exp}=-0.20\pm0.02$
would give a NP signal at the $10\,\sigma$ level, with $\kappa_s\gsim0.2$
from  Eq.~(\ref{phimaxmin}). In analogy to the discussion 
in Subsection~\ref{ssec:comb-Bd}, a determination of  $\kappa_s$
with 10\% uncertainty requires $f_{B_s}\hat B_{B_s}^{1/2}$ with 5\% accuracy, 
i.e.\  the corresponding error in (\ref{HPQCD}) has to be reduced 
by a factor of 2.

Since the discussion given so far does not refer to a specific model of NP, the 
question arises whether there are actually extensions of the SM that still allow 
large CP-violating NP phases in $B^0_s$--$\bar B^0_s$ mixing.

\section{Specific Models of New Physics}\label{sec:NP-models}

In this section, we address the impact of the CDF measurement of
$\Delta M_s$ on
two popular scenarios of NP, to wit
\begin{itemize}
\item an extra $Z'$ boson with flavour non-diagonal couplings;
\item generic effects in the minimal supersymmetric extension of the SM
  (MSSM) in the ``mass insertion approximation''.
\end{itemize}
We would like to stress that our examples for NP
scenarios should be viewed as illustrative rather than
comprehensive and are not intended to compete with more dedicated analyses.

\subsection{\boldmath $Z'$ Gauge Boson with Non-Universal Couplings}

Let us start with the effect of an extra $U(1)'$ gauge boson $Z'$,
which is the most simple 
application  of the model-independent method discussed in
Sections~\ref{sec:Bd} and \ref{sec:Bs}.
The existence of a new  $Z'$ gauge boson can
induce  FCNC processes at tree-level if the $Z'$ coupling
to physical fermions is non-diagonal. Such $Z'$ bosons often occur,
for instance, in the
context of grand unified theories (GUTs), superstring theories, and theories
with large extra dimensions, see, for instance,
Refs.~\cite{Langacker,more-string-inspirations}.
In this paper, we illustrate the constraints on an extra $Z'$ under
the conditions that 
\begin{itemize}
\item the $Z$ couplings stay flavour diagonal, i.e.\
$Z$--$Z'$ mixing is negligible and the $Z$ does not contribute to $B$
mixing; 
\item the $Z'$ has flavour non-diagonal couplings only to
left-handed quarks, which means that its effect is described
by only one complex parameter.
\end{itemize}
 Note that the $Z'$ contribution to $B_s$
mixing is related to that for hadronic, leptonic and semileptonic
decays in specific models where the $Z'$ coupling to light quarks and
leptons is known; in this paper, however, we treat the $Z'$ in a
model-independent way and assume its couplings to the $b_L$ and $s_L$
quark fields as independent. We only discuss the $B_s$-system and
closely follow the notations of Ref.~\cite{barger},
where an earlier analysis of this scenario was given.

A purely left-handed off-diagonal $Z'$ coupling to $b$ and $s$ quarks
gives the following contribution to $M_{12}^s$:\footnote{Strictly
  speaking, $\hat{\eta}^B\hat{B}_{B_s}$ should be taken at LO
  accuracy; here, we effectively absorb the (small) difference between LO and
  NLO expressions into the definition of  $\rho_L$.}
\begin{equation}
M_{12}^{s,Z'} = \frac{G_{\rm F}}{\sqrt{2}}\,\rho_L^2 e^{2i\phi_L}\,\frac{4}{3}\,\hat{\eta}^{B}
\hat{B}_{B_s} f_{B_s}^2 M_{B_s} \,,
\end{equation}
where $\rho_L e^{i\phi_L}\equiv(g' M_Z)/(g M_{Z'})\,B_{sb}^L$ is
defined in terms of the SM $U_Y(1)$ gauge coupling $g$, the $U(1)'$
coupling $g'$, the respective gauge boson masses $M_{Z,Z'}$ and the
FCNC coupling $B_{sb}^L$ of the $Z'$ to $b_L$ and $s_L$. Generically,
one would expect $g/g' = {\cal O}(1)$, if both $U(1)$ groups have the
same origin, for instance in a GUT framework, and
$M_Z/M_{Z'} = {\cal O}(0.1)$ for a TeV-scale $Z'$. If in addition 
the size of the $Z'$ couplings $B^L$ is set by the quarks' Yukawa
couplings, one also expects $|B_{sb}^L| \approx |V_{ts}^*
V_{tb}|$ and $\rho_L = {\cal O}(10^{-3})$.

The impact of the CDF measurement of $\Delta M_s$ on this model can be
directly read off Fig.~\ref{fig:MDs-NP} through the identifications
$$
\rho_L \leftrightarrow (\kappa_s/f)^{1/2},\qquad \phi_L
\leftrightarrow \sigma_s/2\,,$$
with
$$f = \frac{16\pi^2}{\sqrt{2}}\, \frac{1}{G_{\rm F} M_W^2 S_0(x_t)
  |V_{ts}|^2} = (3.57\pm 0.01)\cdot 10^5\,.$$
Presently, values of $\kappa_s$ as large as 2.5 are still allowed, see Fig.~\ref{fig:MDs-NP},
which corresponds to
\begin{equation}\label{47}
\rho_L < 2.6\cdot 10^{-3}\,.
\end{equation}
If a non-zero value of the NP phase $\phi_s^{\rm NP}$ should be
measured at the LHC, this value can be immediately translated into a
lower bound on $\rho_L$, using (\ref{phimaxmin}). Assuming
$\phi_s^{\rm NP} = -10^\circ$, one has
\begin{equation}\label{48}
\sin\phi_s = -0.2 \leftrightarrow \rho_L > 0.5\cdot 10^{-3}\,,
\end{equation}
and $\kappa_s< 0.5\leftrightarrow \rho_L < 1.2\cdot 10^{-3}$. 
Any more precise constraint on $\rho_L$ will depend on the progress in
lattice determinations of $f_{B_s}\hat{B}_{B_s}^{1/2}$. 

The upper bound on $\rho_L$ given in Eq.~(\ref{47}) can be converted
  into a lower bound on the $Z'$ mass:
\begin{equation}
1.5\,{\rm TeV}\left(\frac{g'}{g}\right)
  \left|\frac{B_{sb}^L}{V_{ts}}\right| < M_{Z'}\,.
\end{equation}
In the scenario of (\ref{48}), there is
also an upper bound and the lower bound is raised:
\begin{equation}\label{zprimebound}
3\,{\rm TeV}\left(\frac{g'}{g}\right)
  \left|\frac{B_{sb}^L}{V_{ts}}\right| < M_{Z'} < 
7.5\,{\rm TeV}\left(\frac{g'}{g}\right)
  \left|\frac{B_{sb}^L}{V_{ts}}\right|.
\end{equation}
We would like to stress again that these bounds apply to a model where
the $Z'$ has flavour non-diagonal couplings only to left-handed quarks.
Eq.~(\ref{zprimebound}) can be compared to the existing lower bounds
on the $Z'$ mass from direct
searches, as for instance quoted by CDF \cite{CDFZprime}; these limits
are model-dependent, but in the ballpark of $\sim 800\,$GeV, which is 
perfectly compatible with (\ref{zprimebound}). On the other
hand, if a $Z'$ was found in direct searches at the Tevatron or the
LHC, the bounds on $\rho_L$ would constrain its couplings. This is
particularly interesting in a framework with nearly family-universal
couplings and illustrates the potential synergy between direct
searches for NP and constraints from flavour physics.

Note that (\ref{47}) can also be translated into an upper bound on the
branching ratio of $B_s\to\mu^+\mu^-$, at least if the coupling of
the $Z'$ to $\mu^+\mu^-$ is known. The relevance of such a bound is not
quite clear, however, since we have set the coupling of the $Z'$ to
right-handed fermions to 0.

\subsection{\boldmath MSSM in the Mass Insertion Approximation}

Let us now discuss $B$ mixing in supersymmetry. Whereas in the SM
flavour violation is parametrized by the CKM matrix, in SUSY there are
many more possible ways in which both lepton and quark flavours can
change. This is because scalar quarks and leptons carry the flavour
quantum numbers of their SUSY partners, which implies that flavour
violation in the scalar sector can lead to flavour violation in the
observed fermionic sector of the theory. The parameters controlling
flavour violation in the MSSM are quite numerous -- there are about 
100 soft SUSY breaking parameters which could give rise to huge --
and unobserved -- flavour violation. One way to defuse this so-called
SUSY flavour problem is to assume that the squark (and slepton)
masses are approximately aligned with the quark (and lepton)
masses. ``Alignment'' means that, in the basis of physical states,
where the fermion masses are diagonal, the scalar mass matrices 
are approximately diagonal as well. In
this case, one can treat the off-diagonal terms in the sfermion mass
matrices,
$$
(\delta^f_{ij})_{AB} \equiv (\Delta m^2_{ij})_{AB}/m_{\tilde f}^2\,,
$$
as perturbations.
Here $i,j=1,2,3$ are family indices, $A,B=L,R$ refers to the
``chirality'' of the sfermions\footnote{Sfermions are scalar particles and
  hence have no chirality; the labels $L$ and $R$ refer to the fact
  that they are the SUSY partners of left- and right-handed quark
  fields, respectively.} and $m_{\tilde f}$ is the average
sfermion mass. This so-called mass insertion
approximation (MIA) has been first introduced in Ref.~\cite{Raby}, and was
extensively applied to FCNC and CP-violating phenomena in
Ref.~\cite{masiero}. Its strength is the fact that it is
independent of specific model assumptions on the values of soft
SUSY-breaking parameters, but its weakness is that there are many free
parameters, so there is a certain loss of predictive power.
In this paper, we do not attempt a sophisticated analysis,
which will only be possible once a full NLO calculation of the
corresponding short-distance functions has become
available, which is in preparation, see Ref.~\cite{MIA_NLO}.
Rather, we would like to illustrate the impact of the constraints from
$\Delta M_s$ on the dominant mass insertions, along the lines of, for
instance, Refs.~\cite{emi,EM}. Bounds on mass insertions from $B_d$
mixing have been investigated in Ref.~\cite{BdMIA}.

In supersymmetric theories the effective Hamiltonian
${\cal H}_{\mathrm{eff}}^{\Delta B=2}$ responsible for $B$ mixing, see
Eq.~(\ref{eq1}), is generated by the SM box diagrams
with $W$ exchange and box diagrams mediated by 
charged Higgs, neutralino, photino, gluino and chargino exchange. For small
values of $\tan\beta_{\rm SUSY}$, which is the ratio of vacuum
expectation values of the two MSSM Higgs doublets,
the Higgs contributions are suppressed by the quark masses and can be
neglected.
Photino and neutralino diagrams are also heavily suppressed compared
to those from gluino and chargino exchange, due to the smallness of
the electroweak couplings compared to $\alpha_s$. The
gluino contribution was calculated in Ref.~\cite{masiero}, the
chargino one in Ref.~\cite{GK}. The analysis of Ref.~\cite{emi} has
shown that the chargino contributions are also very small, so that the
$B^0_s$--$\bar B^0_s$ transition matrix element is given, to good accuracy, by
\begin{equation}
M^s_{12} = M_{12}^{s,\mathrm{SM}} +  M_{12}^{s,\tilde{g}}\,,
\end{equation}
where $M_{12}^{s,\mathrm{SM}}$ and $M_{12}^{s,\tilde{g}}$ 
indicate the SM and gluino contributions, respectively. 
It turns out that the largest
contribution to $M_{12}^{s,\tilde{g}}$ comes from terms in
$(\delta^d_{23})_{LL} (\delta^d_{23})_{RR}$, whereas chirality
flipping $LR$ and $RL$ mass insertions are only poorly constrained
from $B_s$ mixing, but dominantly enter $b\to s\gamma$ decays. The
bounds on $(\delta^d_{23})_{LR}$ and $(\delta^d_{23})_{RL}$ posed by
the corresponding branching ratio have been
investigated in Ref.~\cite{marco1}, a recent update can be found in Ref.~\cite{luca}.
As for the chirality-conserving mass insertions, the impact of the D0
bound (\ref{D0-bound}) has been studied in
Refs.~\cite{CS,EM}. Here we set all but one mass insertion to 0 and 
restrict ourselves to bounds on $(\delta^d_{23})_{LL}$ 
and the impact of a future measurement of $\phi_s$ on these bounds. 

\begin{figure}
$$\epsfxsize=0.47\textwidth\epsffile{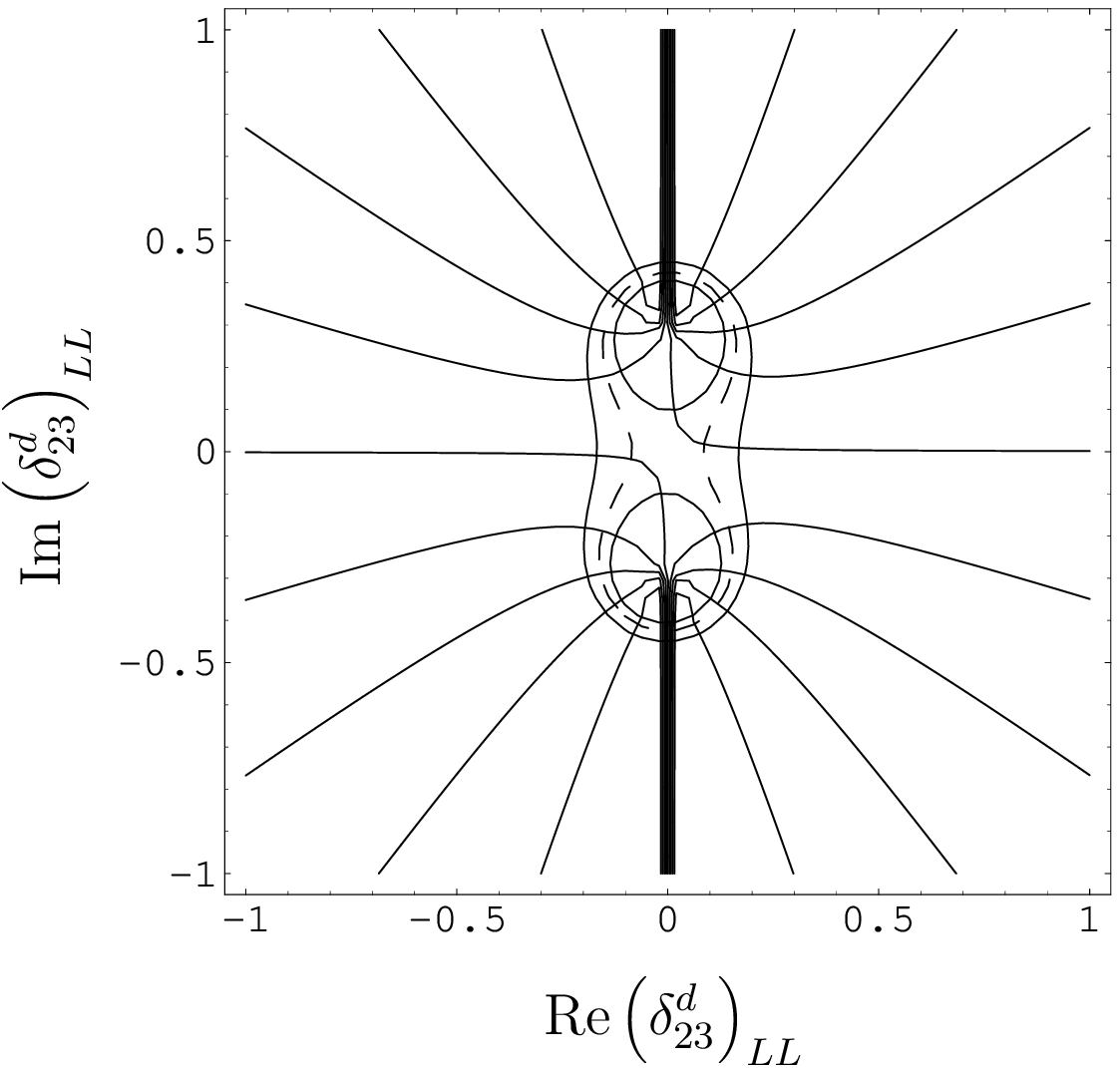}
\qquad\epsfxsize=0.47\textwidth\epsffile{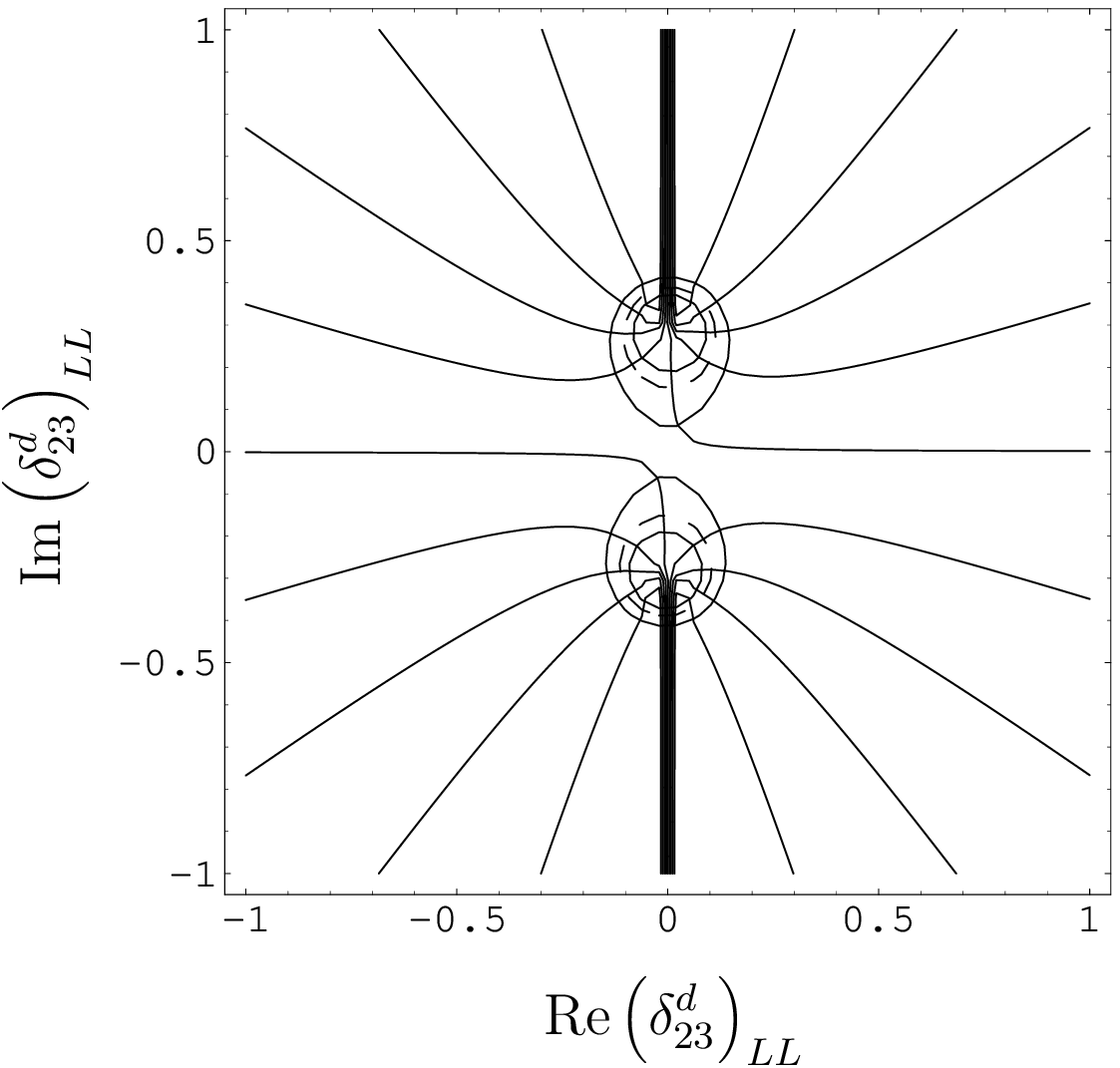}$$
\vspace*{-1cm}
\caption[]{1$\,\sigma$ constraints on $(\delta_{23}^d)_{LL}$  from $\Delta M_s$
  (closed lines). Left panel: JLQCD results (\ref{JLQCD}), right
  panel: (HP+JL)QCD results (\ref{HPQCD}). The open lines denote constraints posed
  by a measurement of $\phi_s$: the curves in the upper right and
  lower left quadrant correspond, from bottom to top in the upper
  quadrant, and
  top to bottom in the lower quadrant, 
  to $\phi_s\in\{0^\circ,36^\circ,72^\circ,108^\circ,144^\circ\}$, whereas those in the
  upper left and lower right quadrant correspond to phases between
  $-180^\circ$ and $0^\circ$.}\label{fig:MIA}
\end{figure}

The effective $\Delta B=2$ Hamiltonian in the MSSM contains a total of eight
operators as compared to only one in the SM. The corresponding
hadronic matrix elements (bag parameters) have been calculated, in
quenched approximation, in Ref.~\cite{lat2}. The evolution of the
Wilson coefficients from $M_{\rm S}$, the scale where the SUSY
particles are integrated out, to $m_b$  is known to next-to-leading order
\cite{BuJa,BdMIA}. The expression for $\Delta M_s$ in the MSSM then
depends on $M_{\rm S}$, $m_{\tilde q}$, the average sfermion mass, and
$m_{\tilde g}$, the gluonino mass. We take $m_{\tilde q} =
500\,{\rm GeV} = m_{\tilde g}$ and also $M_{\rm S} = 500\,{\rm
  GeV}$ as illustrative values. 
We then obtain the constraints on ${\rm Re}\,(\delta^d_{23})_{LL}$ and ${\rm
  Im}\,(\delta^d_{23})_{LL}$ shown in Fig.~\ref{fig:MIA}.
The closed curves in the centre of the plots correspond to the allowed
values of the real and imaginary part of $(\delta_{23}^d)_{LL}$ after
the measurement of $\Delta M_s$; note that the experimental value of
  $\Delta M_s$ is incompatible with the SM prediction at 1.5$\,\sigma$
  level when the (HP+JL)QCD lattice data are used, Eq.~(\ref{36}), hence the origin is
  excluded in the right panel. The open lines correspond to constraints
  imposed by a measurement of the mixing phase $\phi_s$, as explained
  in the caption. It is obvious that at present no value of $\phi_s$
  is excluded and that the precise measurement of the mixing phase,
  expected to take place at the LHC, 
will considerably restrict the parameter space
  of SUSY mass insertions. 

If SUSY is found at the LHC, and the gluino and average squark masses
are measured, the results from MIA analyses of flavour processes will help
to constrain the soft SUSY breaking terms and hence the -- yet to be
understood -- mechanism of SUSY breaking. Given the sheer number of
these terms (about 100), it will be very difficult to resolve the
richness of SUSY breaking  from direct SUSY searches alone, which 
will have to be complemented
by constraints (or measurements) from flavour physics -- which, in turn, will become
more expressive, once the direct searches will have provided the
relevant mass scales. 

\section{Conclusions and Outlook}\label{sec:concl}
The FCNC processes of $B^0_d$--$\bar B^0_d$ and $B^0_s$--$\bar B^0_s$
mixing offer interesting probes to search for signals of physics beyond the SM.
Although the former phenomenon is well established since many years, the latter 
has only just been observed at the Tevatron, thereby raising in particular the question 
of the implications for the parameter space of NP. 

The current situation can be summarized as follows: the experimental value
of the mass difference $\Delta M_d$ and the recently measured $\Delta M_s$
agree with the SM. However, the SM predictions of these quantities suffer
from large uncertainties. In particular, some lattice calculations 
((HP+JL)QCD) indicate a value of $\Delta M_s^{\rm SM}$ that is 1.5$\,\sigma$ 
larger than the experimental CDF value, whereas the JLQCD results show no 
such effect. A similar pattern arises at the 1$\,\sigma$ level in the $B_d$-meson
system. In view of these uncertainties, values of $\kappa_{d,s}$, the
strength of the NP contributions to $B_{d,s}$ mixing, as large
as 2.5 are still allowed by the experimental values of  $\Delta M_{d,s}$,
and the new CP-violating phases $\sigma_{d,s}$ are essentially unconstrained.
Complementary information is provided by CP violation. Interestingly, the 
impressive measurement of mixing-induced CP violation in $B^0_d\to J/\psi K_{\rm S}$ 
(and similar modes) at the $B$ factories may indicate a small -- 
but noticeable -- CP-violating NP phase $\phi_d^{\rm NP}$ around $-10^\circ$,
which would have a drastic impact on the allowed region in the 
$\sigma_{d}$--$\kappa_{d}$ plane and would result in a lower bound on
$\kappa_d$ of $\approx 0.2$. In any case, the experimentally excluded 
large values of $\phi_d^{\rm NP}$ reduce the upper bound $\kappa_d\approx2.5$ 
significantly to $0.5$. On the other hand, no information about 
$\phi_s^{\rm NP}$ is currently available, so that we are left with the
large range of $0\lsim\kappa_s\lsim2.5$.

The following quantities play a key r\^ole for these studies:
mixing:
\begin{itemize}
\item The CKM parameters $\gamma$ and $R_b\propto |V_{ub}/V_{cb}|$,
which enter the analysis of $B^0_d$--$\bar B^0_d$ mixing in a complementary 
manner. Whereas the UT angle $\gamma$ is currently a significant source of 
uncertainty for the SM prediction of $\Delta M_d$ (and $\Delta M_s/\Delta M_d$), 
$R_b$ is crucial for the detection of a NP phase $\phi_d^{\rm NP}$. Thanks to the 
LHCb experiment, the situation for $\gamma$ will improve dramatically in the future, 
where we assumed $\gamma=(70\pm5)^\circ$ in our 2010 benchmark scenario.
Concerning $R_b$, the error of $|V_{cb}|$ has already a marginal impact.
However, there is currently a $1\,\sigma$ discrepancy between the 
inclusive and exclusive determinations of $|V_{ub}|$, pointing towards 
$\phi_d^{\rm NP}\approx-10^\circ$ and $\phi_d^{\rm NP}\approx0^\circ$, respectively.
Consequently, it is crucial to clarify this situation and to reduce the uncertainty 
of $|V_{ub}|$. In our benchmark scenario, we assume that the central 
value of $|V_{ub}|_{\rm incl}$  will be confirmed, and that its uncertainty shrinks
to $5\%$ due to experimental and theoretical progress. It is an advantage of
the $B_s$-meson system that the SM analysis of  its mixing parameters is essentially
unaffected by CKM uncertainties. 

\item The hadronic parameters $f_{B_q}\hat{B}_{B_q}^{1/2}$, which enter the
SM predictions of  $\Delta M_q$. For a determination of $\kappa_q$
with 10\% uncertainty, the errors of the (HP+JL)QCD lattice results have
to be reduced by a factor of 2. The hadronic uncertainties are smaller if one 
considers the ratio $\Delta M_s/\Delta M_d$, involving the $SU(3)$-breaking 
parameter $\xi$. Presently, there is no indication of this ratio to deviate 
from its SM prediction, but there is still a large uncertainty. In our
2010 benchmark scenario, the error from $\xi$ would match that from $\gamma$. Nevertheless,
it will probably be challenging to detect NP through deviations of $\rho_s/\rho_d$
from 1. Moreover, a result in agreement with 1 does not allow any conclusion 
about the presence or absence of NP, as $\rho_s$ and $\rho_d$ may both
deviate similarly from 1, except for excluding certain NP scenarios,
like for instance Higgs penguins enhanced by large values of $\tan\beta_{\rm SUSY}$.
\end{itemize}

Concerning the prospects for the search for NP through $B^0_s$--$\bar B^0_s$
mixing at the LHC, it will be very challenging if essentially no CP-violating effects
will be found in $B^0_s\to J/\psi \phi$ (and similar decays). On the other hand,
as we demonstrated in our analysis, even a small phase 
$\phi_s^{\rm NP}\approx-10^\circ$ (inspired by the $B_d$ data) would lead
to CP asymmetries at the $-20\%$ level, which could be unambiguously detected 
after a couple of years of data taking, and would not be affected by
hadronic uncertainties. Conversely, the measurement of such an asymmetry would 
allow one to establish lower bounds on the strength of NP contribution -- even if 
hadronic uncertainties still preclude a direct extraction of this contribution from 
$\Delta M_s$ -- and to dramatically reduce the allowed region in the NP parameter 
space. In fact, the situation may be even more promising, as specific scenarios of NP 
still allow large new phases in $B^0_s$--$\bar B^0_s$ mixing, also after the 
measurement of $\Delta M_s$. We have illustrated this exciting feature by 
considering models with an extra $Z'$ boson and SUSY scenarios with an 
approximate alignment of quark and squark masses.

In essence, the lesson to be learnt from the CDF measurement of
$\Delta M_s$ is that NP may actually be hiding in $B^0_s$--$\bar B^0_s$
mixing, but is still obscured by parameter uncertainties, some of which will 
be reduced by improved statistics at the LHC, whereas others require dedicated 
work of, in particular, lattice theorists. The smoking gun for the presence of NP 
in $B^0_s$--$\bar B^0_s$ mixing will be the detection of a non-vanishing 
value of $\phi_s^{\rm NP}$ through CP violation in $B^0_s\to J/\psi\phi$. Let us
finally emphasize that the current $B$-factory data may show -- in addition 
to $\phi_d^{\rm NP}\approx-10^\circ$ -- other first indications of new sources of
CP violation through measurements of $B^0_d\to\phi K_{\rm S}$ and $B\to\pi K$
decays, which may point towards a modified electroweak penguin sector. 
All these examples are yet another demonstration that flavour physics is
not an optional extra, but an indispensable
ingredient in the pursuit of NP, also and in particular in the era of the LHC.

\section*{Acknowledgements}

We would like to thank Martin L\"uscher for a discussion of the current
status of lattice calculations of $B$ mixing parameters.

\end{document}